\documentclass[journal]{IEEEtran}
\usepackage{commath,graphicx,afterpage,subfigure,amsmath,amsfonts,amssymb,algorithm,algorithmic}
\usepackage{cite,hyperref,multirow,bm,multicol,booktabs,threeparttable,extpfeil}

\DeclareMathOperator*{\argmax}{arg\,max}

\begin{document}

\title{Multimodal Brain-Computer Interfaces:\\ AI-powered Decoding Methodologies}

\author{Siyang~Li, Hongbin~Wang, Xiaoqing~Chen, and Dongrui~Wu, \IEEEmembership{Fellow,~IEEE}
\thanks{This research was supported by the Open Foundation of Henan Key Laboratory of Brain Science and Brain-Computer Interface Technology under grant HNBBL230204.}
\thanks{S.~Li, H.~Wang, X.~Chen, and D.~Wu are with the Ministry of Education Key Laboratory of Image Processing and Intelligent Control, School of Artificial Intelligence and Automation, Huazhong University of Science and Technology, Wuhan 430074, China. S.~Li is also with the Henan Key Laboratory of Brain Science and Brain Computer Interface Technology, School of Electrical and Information Engineering, Zhengzhou University.}
\thanks{Corresponding Authors: Dongrui Wu (drwu09@gmail.com).}}

\markboth{Under Review at IEEE Reviews in Biomedical Engineering}
{Li \MakeLowercase{Li~\emph{et al.}}: }
\maketitle

\begin{abstract}
Brain-computer interfaces (BCIs) enable direct communication between the brain and external devices. This review highlights the core decoding algorithms that enable multimodal BCIs, including a dissection of the elements, a unified view of diversified approaches, and a comprehensive analysis of the present state of the field. We emphasize algorithmic advancements in cross-modality mapping, sequential modeling, besides classic multi-modality fusion, illustrating how these novel AI approaches enhance decoding of brain data. The current literature of BCI applications on visual, speech, and affective decoding are comprehensively explored. Looking forward, we draw attention on the impact of emerging architectures like multimodal Transformers, and discuss challenges such as brain data heterogeneity and common errors. This review also serves as a bridge in this interdisciplinary field for experts with neuroscience background and experts that study AI, aiming to provide a comprehensive understanding for AI-powered multimodal BCIs.
\end{abstract}


\section{Introduction}

Perception is the process by which sensory information is interpreted to comprehend the surrounding environment \cite{Schacter2009}. A \textit{modality}, therefore, not only refers to how something occurs but also to the way it is experienced or perceived \cite{Baltrusaitis2019}. Perception involves the transmission of signals through the nervous system, which are triggered by physical or chemical stimulation of the sensory system. For instance, visual signals from the retina allow us to see a puppy scampering across the lawn; auditory nerves process sound waves to let us hear a baby's cry; and olfactory receptors detect odor molecules, enabling us to smell the aroma of roasted chicken. Understanding the brain, therefore, is as crucial as understanding these individual modalities, as it serves as the foundation for our interaction with the environment.

\begin{figure*}[htpb] \centering
\includegraphics[width=\linewidth,clip]{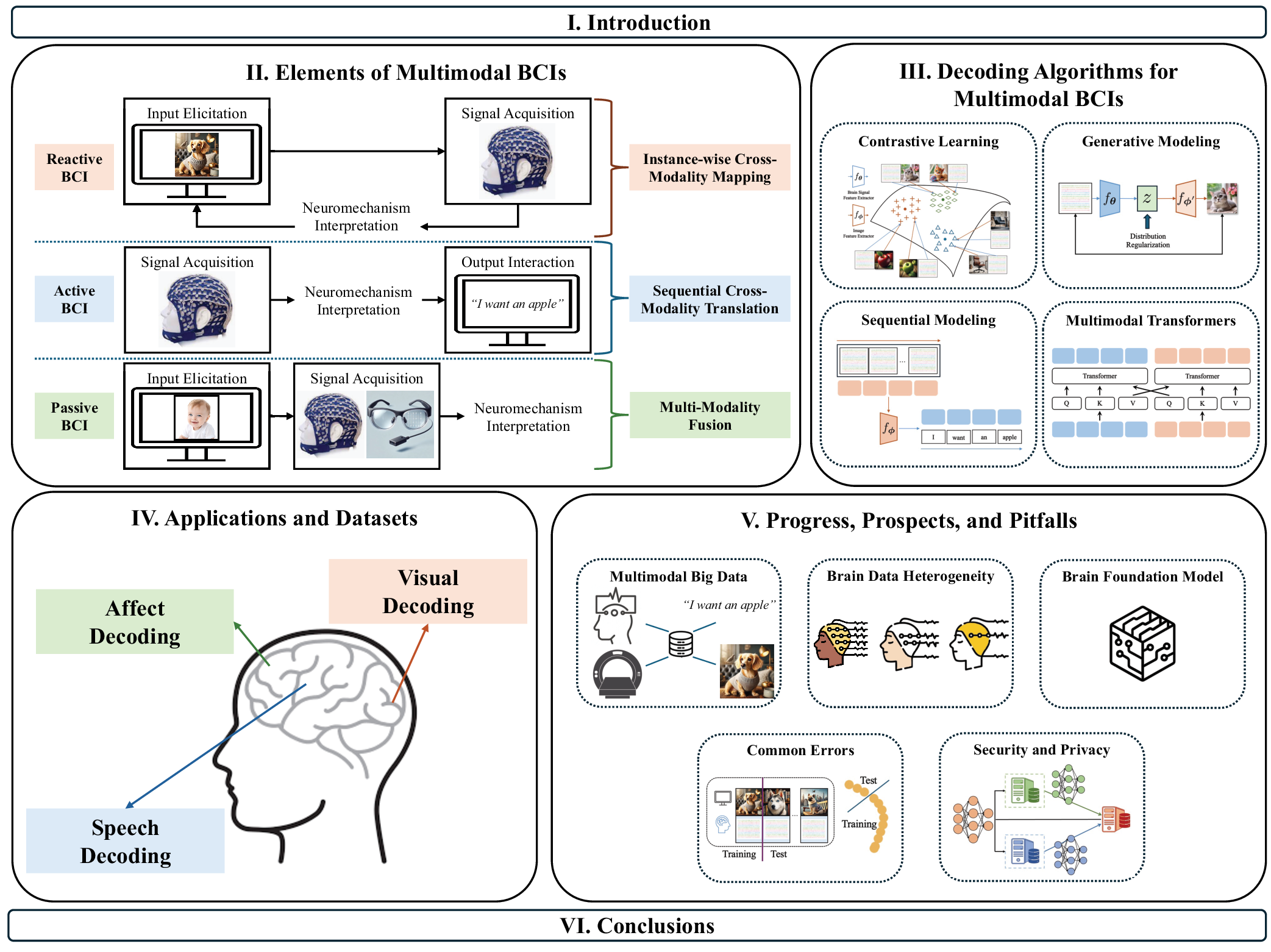}
\caption{The outline of this review.} \label{fig:outline}
\end{figure*}

The brain's ability to interpret and integrate sensory information not only shapes our immediate experiences but also influences our memories, decisions, and emotions. By \textit{decoding} the brain, we can build brain-computer interface (BCI) technologies that offer unprecedented opportunities to bridge the gap between human cognitive functions and machines, potentially revolutionizing how we interact with and manipulate our environment.

This review emphasizes the significance of decoding approaches in \textit{multimodal} BCIs, specifically focusing on the algorithmic perspective enabled by the rapid advancements in artificial intelligence (AI). The seminal work by Gürkök and Nijholt \cite{Guerkoek2012} established foundational principles for multimodal BCIs. With over a decade of advancement in AI, however, an updated analysis of current methodologies and applications is necessary.

Recent breakthroughs in AI decoding algorithms make BCIs highly effective tools for interpreting human cognitive states and intentions. The applications of these technologies span a broad spectrum, including, but not limited to, visual, speech, and affective decoding. By leveraging AI's unique capabilities, these systems can create intuitive interfaces that improve the quality and accessibility of human-machine interactions. Their societal impact is particularly profound in areas such as healthcare and rehabilitation, offering new possibilities for enhancing human capabilities.

To emphasize, the classic definition of multimodality has primarily focused on multiple types of inputs, with fusion tasks being the central problem addressed. This review, however, extends the scope to include additional perspectives beyond multimodal fusion. Specifically, we cover instance-wise cross-modality mapping and sequential cross-modality translation tasks. The goal is to provide a deeper and more comprehensive understanding of the interaction between modalities, with a particular focus on decoding algorithms in the context of multimodal BCIs.

The rest of this paper is organized as follows: Section~\ref{sect:background} reviews related work and the background of BCIs. Section~\ref{sect:background} details the components of BCIs, categorizes multimodal tasks, and describes the types of brain data. Section~\ref{sect:applications} presents representative multimodal BCI applications, corresponding public datasets, and their core decoding algorithms. Section~\ref{sect:prospects} discusses future research prospects and common pitfalls. Finally, Section~\ref{sect:conclusions} concludes the review. An outline of the paper is shown in Fig.~\ref{fig:outline}.

\section{Elements of Multimodal BCIs} \label{sect:background}

\subsection{Components of BCIs} \label{sect:bcicomponents}

Multimodal BCIs include four primary components: input elicitation, brain data acquisition, neuromechanism interpretation, and output interaction. Each component plays a critical role in evoking, capturing, and translating neural activity into functional outputs, bridging human cognition and external devices.

\textbf{Input Elicitation} Stimuli are used to elicit specific neural responses from the user. These stimuli can be visual, auditory, or sensory, designed to provoke brain activity that the BCI system can detect and interpret. Stimulus design must prioritize specificity to reduce user confusion, ensure user comfort and safety by avoiding sensory overload or fatigue, and adapt to individual differences in sensory perception and cognitive processing.

\textbf{Brain Data Acquisition} (also known as brain signal acquisition or brain imaging) At the core of any BCI system is its capacity to accurately capture brain activity. This involves measuring electrical activity, magnetic fields, or blood flow changes associated with neural activity using specialized devices. The choice of acquisition method directly impacts the resolution, precision, and potential applications of the BCI, ranging from basic communication aids to complex prosthetic control systems.

\textbf{Neuromechanism Interpretation} Decoding neural activity is essential for effective BCI functionality. This step involves interpreting the brain's responses to stimuli and understanding the neuromechanisms behind them, considering various factors in the processing techniques. Decoding algorithms are further influenced by specific neuro mechanistic characteristics, such as the latency of task-related responses, the temporal window of interest, and the frequency bands retained for analysis. The effectiveness thus highly depends on the precision of algorithms that decode neural recordings.

\textbf{Output Interaction} The final component involves translating interpreted brain signals into actionable outputs. While passive BCIs may not include this step, active BCIs utilize it to enable functions such as synthesizing speech, executing physical commands, or interacting with virtual environments.

These components form the foundation of BCI systems. In the following sections, we explore how AI-powered decoding distinguishes multimodal BCIs from classical designs, enabling sophisticated integration across multiple modalities.

\subsection{Categorization of Multimodal BCIs} \label{sect:perspective}

Multimodal learning has become a prominent research area in AI \cite{Baltrusaitis2019}. With the development of advanced methodologies such as contrastive learning, generative modeling, sequential modeling, multimodal Transformers, etc., the categorization of multimodal BCIs has evolved beyond classic feature/decision-level fusion tasks. This subsection offers a brief peak of the core mechanisms of multimodal interactions, grounded in the classical categorization of BCIs.

\begin{figure*}[htpb] \centering
\includegraphics[width=.8\linewidth,clip]{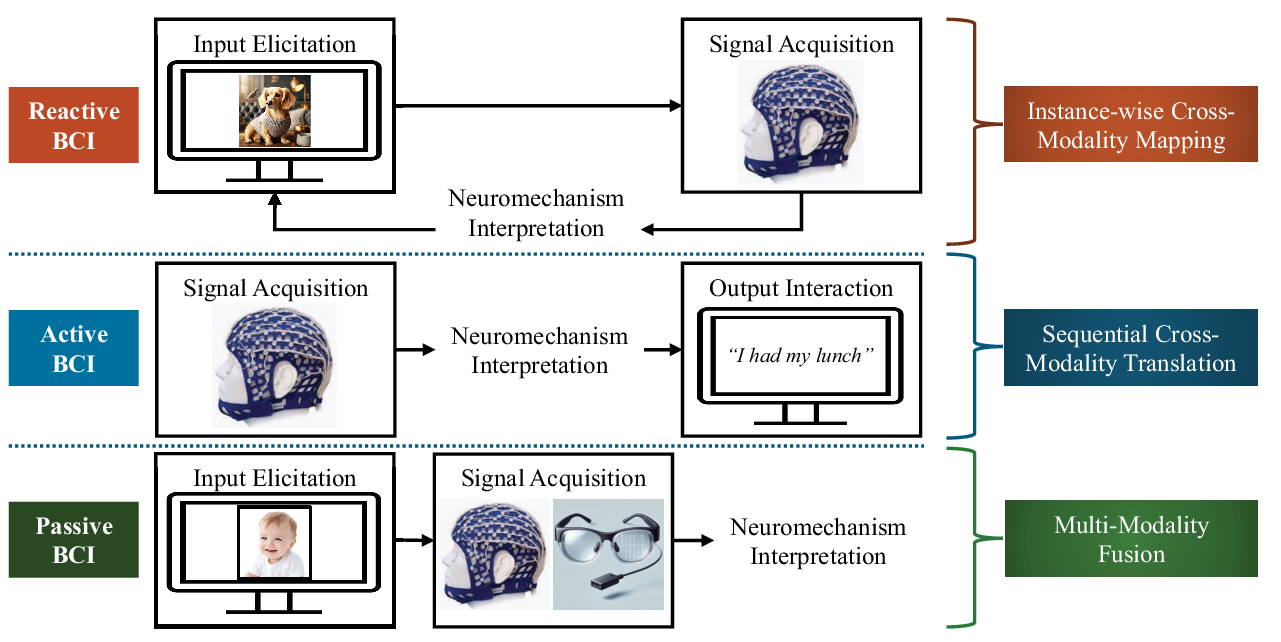}
\caption{Three types of multimodal BCIs and their representative applications. Reactive BCIs involve brain activities stimulated by designed inputs, such as visual decoding, where perceived images are decoded instance-wise. Active BCIs facilitate user-driven communication in sequential forms, such as speech decoding. Passive BCIs record brain activities using various sensors for tasks like affective BCIs. The core mechanisms involved are very distinct in the aspect of AI decoding algorithms for such three types.} \label{fig:categorization}
\end{figure*}

There are various categorizations for BCIs. Based on where the input signal is collected, BCIs can be categorized into non-invasive/invasive/partially invasive \cite{Rao2013}. As illustrated in Fig.~\ref{fig:categorization}, we propose a novel algorithmic categorization that distinguishes BCIs into three types. This categorization integrates the classic reactive/active/passive BCI categorization \cite{Zander2009, Guerkoek2012}:
\begin{itemize}
\item \textbf{Reactive BCIs} use unintended changes in cognition by voluntarily focussing on outside stimuli. It is composed of input elicitation, brain signal acquisition, and neuromechanism interpretation.
\item \textbf{Active BCIs} utilize brain activity that directly correlates with intended actions. It is composed of brain signal acquisition, neuromechanism interpretation, and output interaction.
\item \textbf{Passive BCIs} monitor in the background to detect states based on natural or spontaneous brain activities. It is composed of input elicitation, signal acquisition and neuromechanism interpretation.
\end{itemize}

The algorithmic categorization of multimodal BCIs is detailed below:

\textbf{Instance-wise Cross-Modality Mapping} This task focuses on mapping an entity from one modality to a corresponding entity in another. For example, reactive BCIs decode brain activity to reconstruct input stimuli, such as visual experiences.

Algorithmically, the mapping function $f_{\text{mapping}}(\cdot)$ transforms a sample from modality $\mathbf{x}^{A}$ into an entity in modality $\mathbf{x}^{B}$:
\begin{align}
\mathbf{x}^{B} = f_{\text{mapping}}(\mathbf{x}^{A}).
\end{align}

\textbf{Sequential Cross-Modality Translation} Sequential communication requires the continuous translation of input modality sequences into target modality sequences while preserving context, coherence, and integrity. Active BCIs are particularly suited for tasks like speech decoding, enabling fluent communication.

Algorithmically, the translation function $f_{\text{translation}}(\cdot)$ converts sequential inputs from one modality $(\mathbf{x}_{1}^{A}, \mathbf{x}_{2}^{A}, ..., \mathbf{x}_{T}^{A})$ into another modality:
\begin{align}
(\mathbf{x}_{1}^{B}, \mathbf{x}_{2}^{B}, ..., \mathbf{x}_{T}^{B}) = f_{\text{translation}}\left((\mathbf{x}_{1}^{A}, \mathbf{x}_{2}^{A}, ..., \mathbf{x}_{T}^{A})\right).
\end{align}

\textbf{Multi-Modality Fusion} This classic topic involves combining information from multiple modalities to achieve an integrated decoding result. Passive BCIs utilize this approach by fusing multiple modalities, such as EEG and video data, to improve decoding accuracy.

Algorithmically, the fusion function $f_{\text{fusion}}(\cdot, \cdot)$ integrates samples from two or more modalities, $\mathbf{x}^{A}$ and $\mathbf{x}^{B}$, into a single output label $y$:
\begin{align}
y = f_{\text{fusion}}(\mathbf{x}^{A}, \mathbf{x}^{B}),
\end{align}

It is also viable to build a sequential fusion function that further considers temporal information:
\begin{align}
y_t = f_{\text{seqfusion}}\left((\mathbf{x}_{1}^{A}, \mathbf{x}_{2}^{A}, ..., \mathbf{x}_{T}^{A}); (\mathbf{x}_{1}^{B}, \mathbf{x}_{2}^{B}, ..., \mathbf{x}_{T}^{B})\right),
\end{align}

\textbf{Remark} Observe that the technical problems involved in the tasks for the three types of multimodal BCIs are very distinct. Each task presents unique challenges and opportunities, demanding tailored AI algorithms. Addressing these distinctions has catalyzed the development of novel applications in BCIs.

\subsection{Types of Brain Data} \label{sect:characteristics}

Brain data collection is integral to the functionality and applications of BCIs. While various BCI-related surveys provide detailed descriptions, this subsection offers a concise overview of the most widely used brain data collection methods and their characteristics, summarized in Table~\ref{tab:signals}. These methodologies can be broadly categorized into invasive and non-invasive techniques, each with distinct advantages and limitations that influence their suitability for specific BCI applications.

\begin{table*}[htpb]  \center
\caption{Summary of brain data collection devices and their data characteristics.}  \label{tab:signals}
\begin{tabular}{c|c|c|c|c|c|c}
\toprule
Category & Name & Spatial Resolution & Temporal Resolution & Cost & Universality & Inplant Duration / Time for Setup \\
\midrule
\multirow{4}{*}{Invasive} & ECoG & High & High & Expensive & Low & Short-term \\
& SEEG & High & High & Expensive & Low & Short-term \\
& MEA & Very High & Very High & Very Expensive & Low & Medium-term \\
& Flexible Electrodes & High / Very High & High / Very High & Very Expensive & Low & Long-term \\
\midrule
\multirow{4}{*}{Non-Invasive} & EEG & Low & High & Cheap & High & Quick (Dry) / Moderate (Wet) Setup \\
& MEG & Moderate & High & Very Expensive & Low & Moderate Setup \\
& fMRI & High & Low & Very Expensive & Moderate & Moderate Setup \\
& fNIRS & Moderate & Moderate & Moderate & Moderate & Quick Setup \\
\bottomrule
\end{tabular}
\end{table*}

\textbf{Invasive BCIs} Invasive techniques require surgical implantation of electrodes directly into the brain, offering high-resolution signals essential for precise monitoring and intervention. These methods are predominantly used in clinical and research settings for applications such as disease localization and advanced neural interfacing.
\begin{itemize}
\item \textbf{Electrocorticography (ECoG)} Electrodes are placed directly on the exposed brain surface, providing broad area coverage without deep penetration \cite{Leuthardt2004}. ECoG is used for motor cortex mapping and tasks requiring precise, localized monitoring of brain activity.
\item \textbf{Stereoelectroencephalography (SEEG)} Electrodes are implanted into the brain through small burr holes, enabling the recording of electrical activity from deep brain structures \cite{Talairach1966}. SEEG is widely used for epilepsy diagnosis and pre-surgical planning.
\item \textbf{Microelectrode Arrays (MEA)} Arrays of silicon-based electrodes penetrate brain tissue, capturing single-neuron activity with exceptional spatial and temporal resolution. MEAs are instrumental in detailed neural mapping. For example, the Utah array, the first invasive neural interface approved by the U.S. FDA, has evolved to record from over a thousand electrodes in a compact area \cite{Schwartz2004}.
\item \textbf{Flexible Electrodes} Fabricated using soft polymers or advanced techniques \cite{Musk2019, Jiang2022, Tang2023, Wang2023a, Zhao2023, Zhang2023b, Dong2024, LeFloch2024}, these electrodes conform to the brain's surface or penetrate tissue with minimal damage. They offer high biocompatibility, scalability, and long-term stability, reducing chronic immune responses \cite{Lee2022}.
\end{itemize}

They are also loosely referred to as intracranial electroencephalography \cite{Parvizi2018}.

\textbf{Non-invasive BCIs} Non-invasive techniques \cite{Edelman2024} are safer and more accessible, eliminating the need for surgical procedures. They rely on external sensors or imaging devices, making them suitable for a broader range of users, including consumer and clinical applications.
\begin{itemize}
\item \textbf{Electroencephalography (EEG)} EEG \cite{Roy2019, Gu2021} is one of the most popular non-invasive methods due to its affordability and ease of use. It measures electrical activity through scalp electrodes, supporting applications from medical diagnostics to user interface control. Dry electrodes offer quick setups but may be noisier, while wet electrodes provide better signal quality at the expense of longer preparation times. However, EEG signals are limited to low-frequency bands due to attenuation by the skull and scalp \cite{NicolasAlonso2012}.
\item \textbf{Magnetoencephalography (MEG)} MEG \cite{Cohen1972} records magnetic fields generated by neural activity, offering a balance of high temporal and moderate spatial resolution. It is predominantly used in research for time-sensitive brain activity analysis.
\item \textbf{Functional Magnetic Resonance Imaging (fMRI)} fMRI \cite{Sitaram2008, Du2022} detects blood flow changes associated with brain activity, offering excellent spatial resolution. It is invaluable for mapping neural activity across the entire brain but has limited temporal resolution and high operational costs.
\item \textbf{Functional Near-Infrared Spectroscopy (fNIRS)} Using light to measure blood oxygenation, fNIRS \cite{Joebsis1977} provides moderate spatial and temporal resolution. It is portable, less affected by electrical interference, and suitable for practical and research applications.
\end{itemize}

The choice between invasive and non-invasive methods depends on the application's requirements, including resolution, invasiveness, and cost. Invasive methods are preferred for high-precision clinical or research tasks, while non-invasive approaches dominate consumer and general clinical applications due to their safety and convenience.

\section{Decoding Algorithms for Multimodal Learning} \label{sect:methods}

This section introduces key AI techniques fundamental to multimodal BCIs.

\subsection{Cross-Modality Contrastive Learning} \label{sect:contrastive}

Cross-modality contrastive learning focuses on classification and retrieval tasks by aligning paired samples in the feature space while separating unpaired ones. This approach is particularly effective for instance-wise cross-modality mapping, enabling the alignment of representations from two modalities that share identical semantics\footnote{For readers unfamiliar with such terminologies, it is safe to comprehend ``representation" as ``feature extracted by trained neural networks" \cite{Bengio2013}, in the form of a vector. The term ``semantics" refers to ``meaning/class", e.g., a cat image, text of ``cat", and the brain data of a subject viewing a cat, has identical semantics.}. For example, brain data representations corresponding to viewing a cat can be aligned with those of a cat image. This is achieved through the learning objective detailed below.

\textbf{InfoNCE Objective} The InfoNCE objective \cite{Oord2018CPC, Radford2021} (NCE stands for noise contrastive estimation) takes a symmetrical form. Denote feature extractors for the two modalities $A$ and $B$ as $f_{\boldsymbol{\theta}}(\cdot)$ and $f_{\boldsymbol{\phi}}(\cdot)$, parameterized by trainable neural networks for better feature learning with parameters $\boldsymbol{\theta}$ and $\boldsymbol{\phi}$, respectively. Cross-modality contrastive learning aims to optimize their parameters $\boldsymbol{\theta}$ and $\boldsymbol{\phi}$ with paired inputs $\{(\mathbf{x}_{i}^{A}, \mathbf{x}_{i}^{B})\}_{i=1}^{n}$ from two modalities $A$ and $B$ through the following objective:
\begin{align}
& \mathcal{L}_{A \rightarrow B}^{\text{InfoNCE}} \left((\mathbf{x}^{A}, \mathbf{x}^{B}); f_{\boldsymbol{\theta}}; f_{\boldsymbol{\phi}} \right) = \nonumber \\ & \hfill - \frac{1}{n} \sum_{i=1}^{n} \log \frac{\exp \left(f_{\boldsymbol{\theta}}(\mathbf{x}_{i}^{A}) \cdot f_{\boldsymbol{\phi}}(\mathbf{x}_{i}^{B}) / \tau \right)}{\sum_{j=1}^{n} \exp  \left(f_{\boldsymbol{\theta}}(\mathbf{x}_{i}^{A}) \cdot f_{\boldsymbol{\phi}}(\mathbf{x}_{j}^{B}) / \tau \right)},
\label{eq:infonce-oneside}
\end{align}
where $\exp(\cdot)$ calculates the exponential (taking the logarithm after summing exponentiated logits computes log probabilities more accurately while ensuring numerical stability), and $\tau$ is a temperature hyperparameter that controls the density of the learned representation (a lower value of $\tau$ makes the model more sensitive to differences between the more similar and dissimilar pairs \cite{Wang2021}), the dot product is a similarity calculation between two samples and can practically also be replaced with other functions like cosine similarity.

For alignment of two modalities, the complete objective combines bidirectional alignment that $\mathcal{L}_{A + B}^{\text{InfoNCE}} = \mathcal{L}_{A \rightarrow B}^{\text{InfoNCE}} + \mathcal{L}_{B \rightarrow A}^{\text{InfoNCE}}$. This objective pulls anchor-positive pairs closer and pushes anchor-negative pairs farther in the feature space. For cross-modality mapping tasks, the anchor $f_{\boldsymbol{\theta}}(\mathbf{x}_{i}^{A})$ corresponds to modality $A$, with $f_{\boldsymbol{\phi}}(\mathbf{x}_{i}^{B})$ as the positive pair and $f_{\boldsymbol{\phi}}(\mathbf{x}_{j}^{B})$ as the negative pair.

\begin{figure*}[htpb]
\centering
\subfigure[]{\includegraphics[width=.49\linewidth,clip]{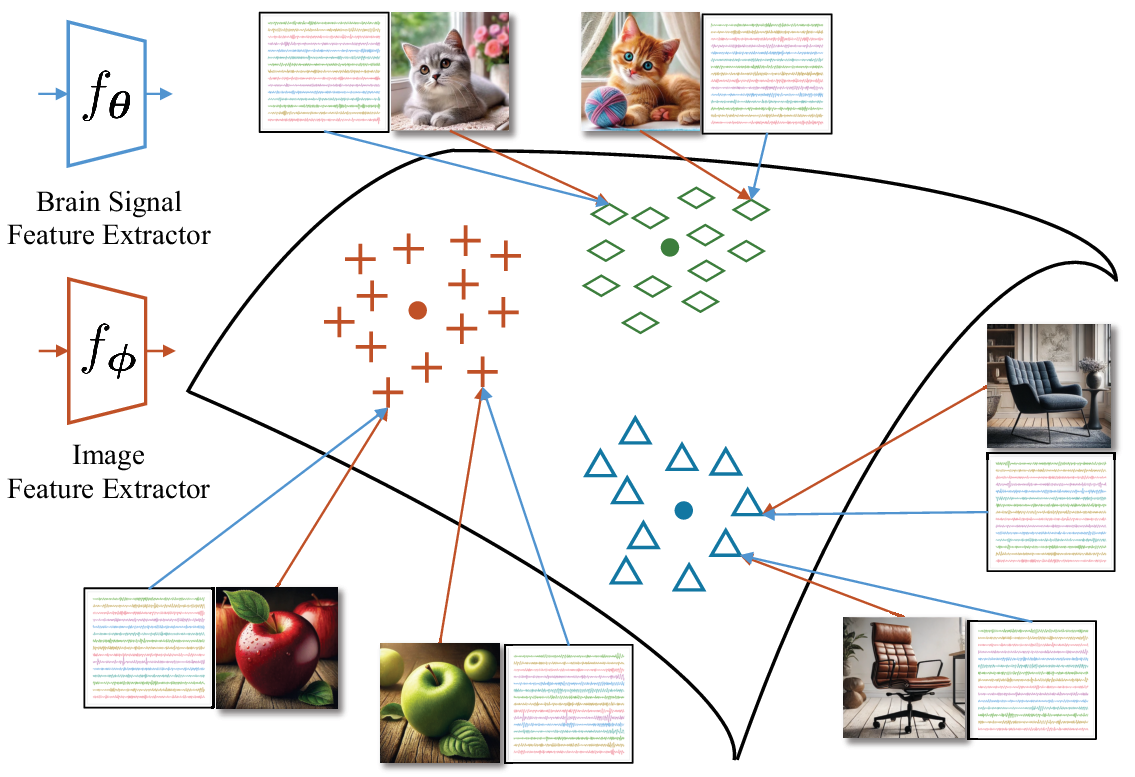}\label{fig:zeroshot_training}}
\subfigure[]{\includegraphics[width=.49\linewidth,clip]{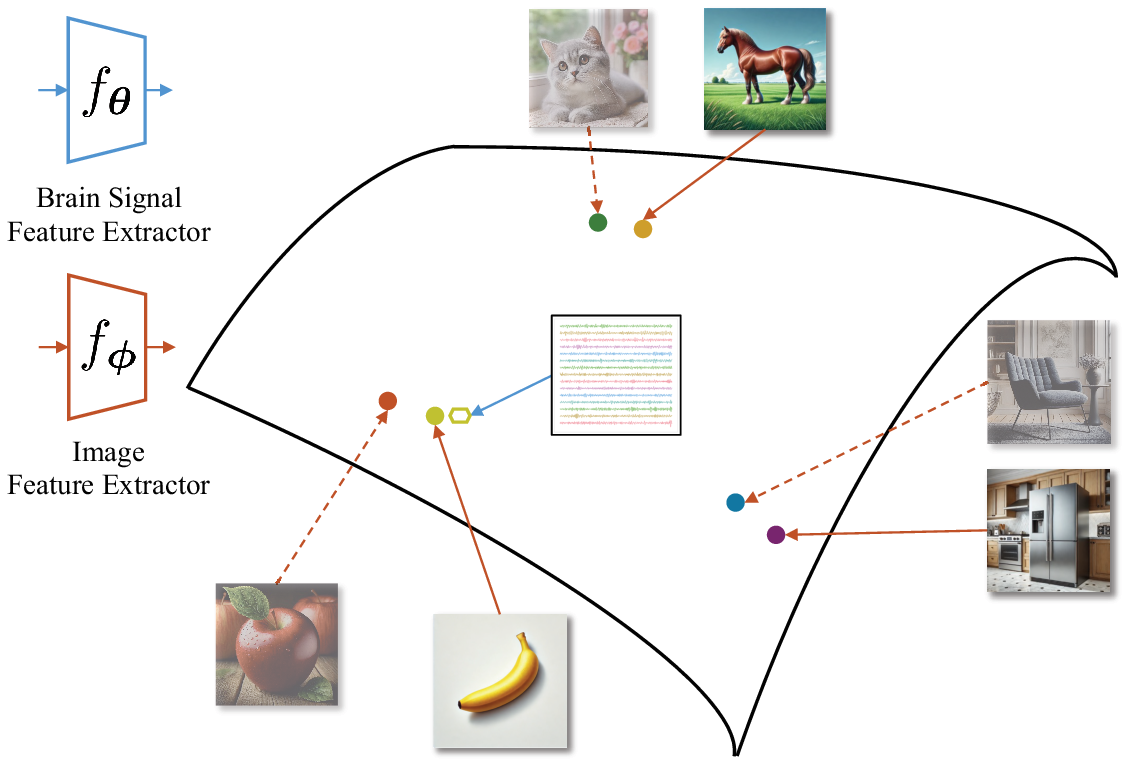}\label{fig:zeroshot_test}}
\caption{Zero-shot classification under cross-modality contrastive learning, using visual decoding as an example. (a) Training stage, where feature extractors for image and brain signal pairings are learned in the latent space using cross-modality contrastive learning. (b) Test stage, where novel classes that were not present in the training set require classification. Retrieval could be performed using similarity metrics in the latent space.} \label{fig:zeroshot}
\end{figure*}

\textbf{Zero-Shot Classification} A notable advantage of cross-modality contrastive learning is its capacity for zero-shot classification \cite{Lampert2009, Socher2013}, which eliminates the dependence on predefined class labels. As illustrated in Fig.~\ref{fig:zeroshot}, zero-shot classification replaces traditional classification with a retrieval objective, enabling the system to generalize to unseen classes during testing.

This is achieved by the InfoNCE objective. It produces feature extractors that generate correlated representations, effectively constructing a mapping function that class prediction can be determined with similarity metrics that
\begin{align}
\mathbf{x}^{B} = f_{\text{mapping}} (\mathbf{x}^{A})= \argmax_{\mathbf{x}_{i}^{B}} \frac{f_{\boldsymbol{\theta}} (\mathbf{x}^{A}) \cdot f_{\boldsymbol{\phi}}{(\mathbf{x}_{i}^{B})}}{\tau}.
\end{align}

\textbf{Extensions and Alternatives} Beyond classic contrastive learning, cross-modality alignment can also incorporate label information \cite{Yang2022, Khosla2020}. Additionally, this framework can be extended to multiple modalities \cite{Mai2022, Lin2022} or replaced with direct alignment techniques \cite{Dou2022, Bao2022}, though these approaches are less common in practice.

\subsection{Cross-Modality Generative Modeling} \label{sect:generative}

\begin{figure*}[htpb]
\centering
\subfigure[]{\includegraphics[width=.4\linewidth,clip]{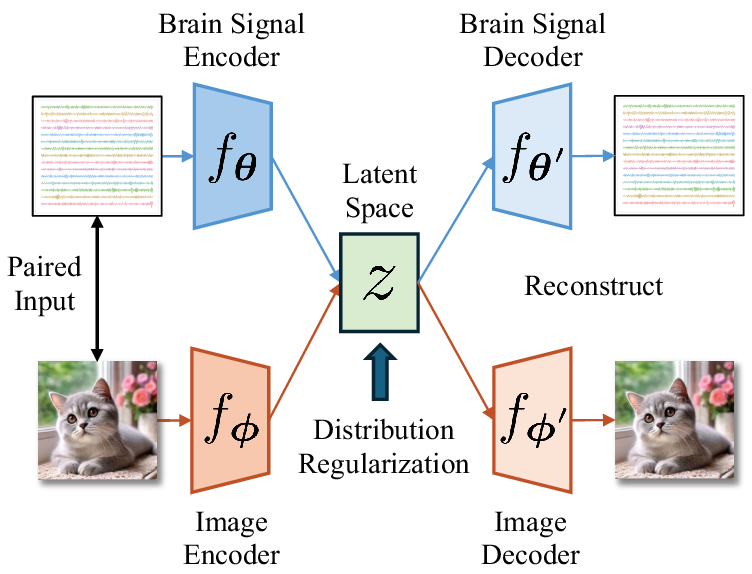}\label{fig:generative_joint}}
\hspace{.1\linewidth}
\subfigure[]{\includegraphics[width=.4\linewidth,clip]{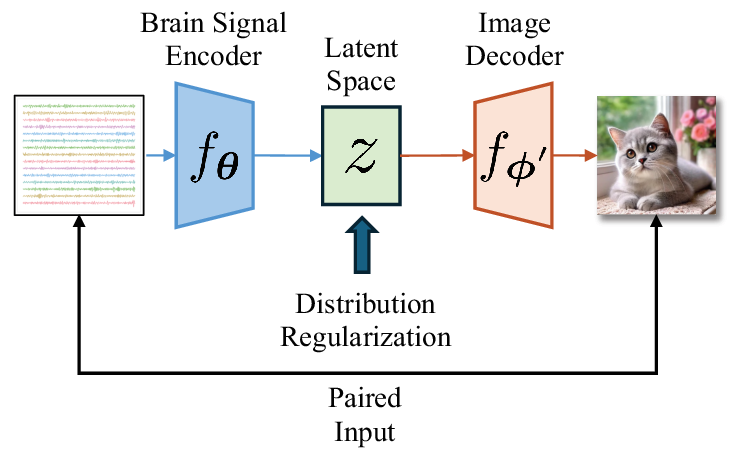}\label{fig:generative_conditional}}
\caption{Two types of cross-modality generative modeling, using two modalities of image and brain signal as an example. (a) Joint latent space, where separate encoders project two modalities into a shared latent space, and separate decoders reconstruct the respective inputs; (b) Conditional latent space, where an asymmetrical encoder and decoder projects inputs from a modality to a latent space and then to another modality.} \label{fig:generative}
\end{figure*}

Generative modeling has become a cornerstone of contemporary AI systems, particularly in enabling cross-modality mapping as summarized in this subsection. Variational generative models, such as variational autoencoders (VAEs) \cite{Kingma2013}, generative adversarial networks (GANs) \cite{Goodfellow2014}, and diffusion models \cite{Ho2020}, achieve this by learning either a joint latent space or a conditional latent space that correlates two modalities. These latent spaces are typically regularized to align with a prior distribution (commonly Gaussian) to facilitate meaningful sampling, interpolation, and generalization. Figure~\ref{fig:generative} illustrates these two paradigms.

\textbf{Joint Latent Space} In joint latent space cross-modality mapping, separate encoders $f_{\boldsymbol{\theta}} (\cdot)$ and $f_{\boldsymbol{\phi}} (\cdot)$ project two modalities into a common space $z$ with reduced dimensionality such that $f_{\boldsymbol{\theta}}(\mathbf{x}^{A}), f_{\boldsymbol{\phi}}(\mathbf{x}^{B}) \in \mathbb{R}^{d}$. Separate decoders $f_{\smash{\boldsymbol{\theta}'}} (\cdot)$ and $f_{\smash{\boldsymbol{\phi}'}} (\cdot)$ reconstruct the respective modalities from latent space back to their original forms. The training process typically employs a Gaussian distribution for regularization:
\begin{align}
& \mathcal{L}_{\text{shared}} \left( (\mathbf{x}^A, \mathbf{x}^B); f_{\boldsymbol{\theta}}; f_{\smash{\boldsymbol{\theta}'}}; f_{\boldsymbol{\phi}}; f_{\smash{\boldsymbol{\phi}'}} \right) = \nonumber \\ & \hfill \mathbb{E}_{q_{\boldsymbol{\theta}}(z | \mathbf{x}^A)} \left[ \log p_{\smash{\boldsymbol{\theta}'}}(\mathbf{x}^A | z) \right] + \mathbb{E}_{q_{\boldsymbol{\phi}}(z | \mathbf{x}^B)} \left[ \log p_{\smash{\boldsymbol{\phi}'}}(\mathbf{x}^B | z) \right] \nonumber \\ & \hfill - D_{\text{KL}} \left( q_{\boldsymbol{\theta}}(z | \mathbf{x}^A) \| p(z) \right) - D_{\text{KL}} \left( q_{\boldsymbol{\phi}}(z | \mathbf{x}^B) \| p(z) \right),
\end{align}
where $q (z | \mathbf{x})$ denotes the approximate posterior distributions of encoders, $p(\mathbf{x} | z)$ the prior likelihood distributions of decoders, $p(z)$ The prior distribution over the latent space which is often assumed to be a standard Gaussian distribution (i.e., $\mathcal{N}(0, I)$), $D_{\text{KL}}(q \| p)$ the Kullback-Leibler (KL) divergence which measures the extent of distribution gap (how much the approximate posterior $q$ diverges from the prior $p$) and regularizes the latent distribution in VAEs.

After training, data from one modality can be mapped to another by encoding it into the shared latent space and decoding it with the target modality's decoder that $\mathbf{x}^{B} = f_{\smash{\boldsymbol{\phi}'}}\left(f_{\boldsymbol{\theta}}(\mathbf{x}^{A})\right)$ or $\mathbf{x}^{A} = f_{\smash{\boldsymbol{\theta}'}}\left(f_{\boldsymbol{\phi}}(\mathbf{x}^{B})\right)$.

\textbf{Conditional Latent Space} In conditional latent space mapping, the representation is explicitly conditioned on the source modality. An encoder $f_{\boldsymbol{\theta}} (\cdot)$ projects input from one modality to a latent space $z$, and a decoder $f_{\smash{\boldsymbol{\phi}'}} (\cdot)$ maps the latent representation to the target modality:
\begin{align}
& \mathcal{L}_{\text{conditional}} \left( (\mathbf{x}^A, \mathbf{x}^B); f_{\boldsymbol{\theta}}, f_{\smash{\boldsymbol{\phi}'}} \right) = \nonumber \\ & \hfill  \mathbb{E}_{q_{\boldsymbol{\theta}}(z | \mathbf{x}^A)} \left[ \log p_{\smash{\boldsymbol{\phi}'}}(\mathbf{x}^B | z) \right] - D_{\text{KL}} \left( q_{\boldsymbol{\theta}}(z | \mathbf{x}^A) \| p(z) \right),
\end{align}
where regularization may involve KL divergence (e.g., in VAEs), progressive diffusion processes (e.g., in diffusion models), or implicit regularization through adversarial training (e.g., in GANs).

Cross-modality mapping in this case follows $\mathbf{x}^{B} = f_{\smash{\boldsymbol{\phi}'}}\left(f_{\boldsymbol{\theta}}(\mathbf{x}^{A})\right)$. Note that such a model is now an asymmetrical, one-way mapping instead.

\subsection{Sequential Modeling} \label{sect:sequential}

\begin{figure*}[htpb] \centering
\includegraphics[width=.92\linewidth,clip]{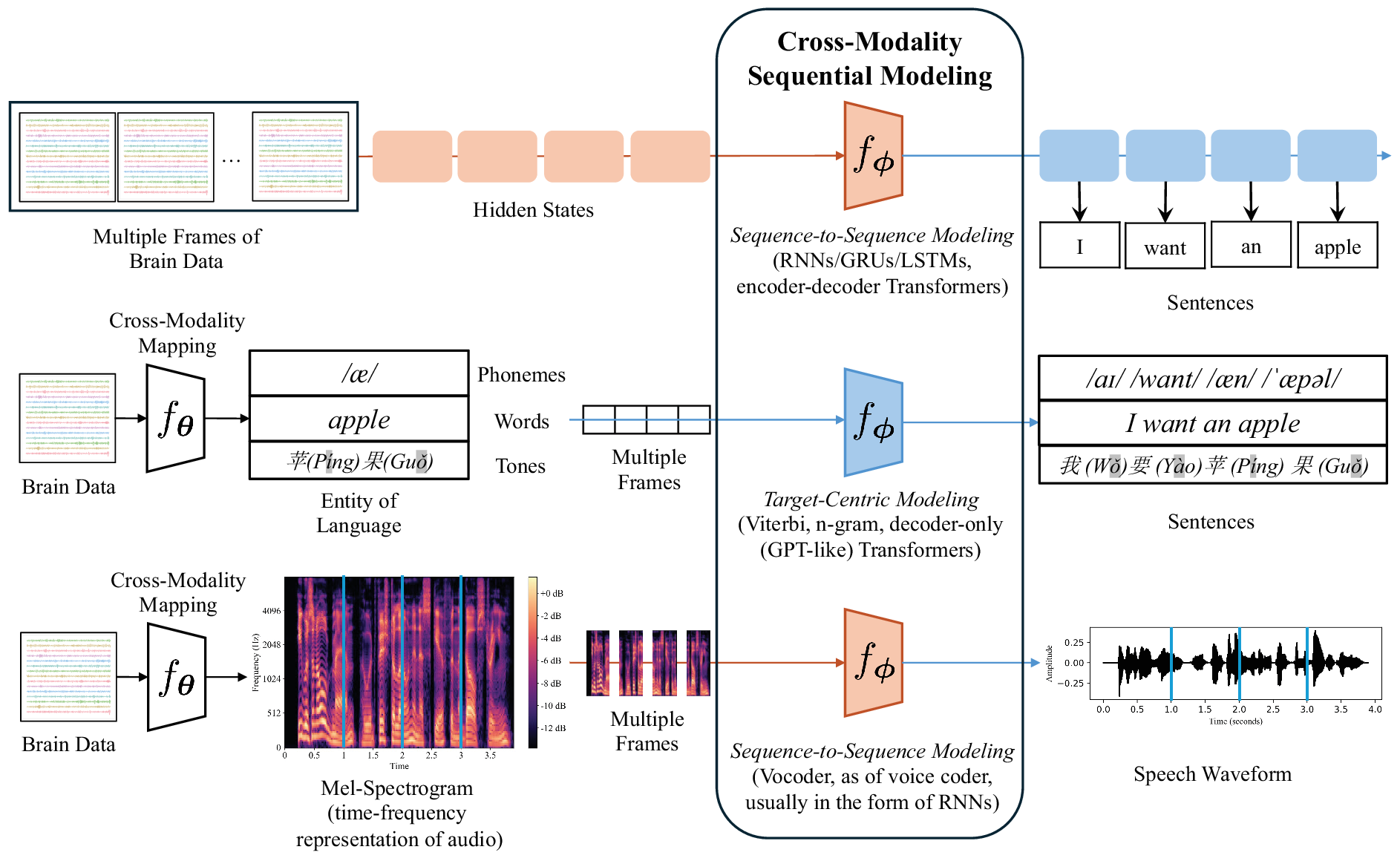}
\caption{Three pipelines for decoding brain recordings into speech and their respective sequential modeling components. The majority of works decode brain data into sentences, either through direct sequence-to-sequence neural networks, or first map to discrete entities of language and then reorganize them. An alternative pipeline directly synthesizes speech waveform, through first mapping to mel-spectrogram, then using vocoder for audio generation.} \label{fig:speech}
\end{figure*}

Sequential modeling has become a fundamental paradigm for analyzing chronologically ordered data, including text, speech, and time series. The aforementioned algorithms only considered learning instance-wise functions across modalities, whereas the sequential nature of such data was ignored. Sequential modeling of brain signal data on the temporal dimension is a more appealing direction compared to instance-wise analysis. Three pipelines for decoding brain data into speech are illustrated in Fig.~\ref{fig:speech}.

Specifically, there are two types of sequential modeling for cross-modality translation. One type focuses on reorganizing the target modality entities to maintain temporal coherence, while the other type considers both the source and target modality entities to improve both temporal coherence and inter-modality alignment.

\subsubsection{Target-Centric Sequential Modeling}

Assuming that a prebuilt cross-modality mapping function $f_{\text{mapping}}$ has already predicted the output probabilities for the target modality $B$, these methods focus on improving fluency, coherence, and sequential consistency of the target modality.

\textbf{Viterbi Algorithm} The Viterbi algorithm finds the most probable sequence of entities in target modality $\mathbf{x}^{B} = (\mathbf{x}_1^{B}, \mathbf{x}_2^{B}, \ldots, \mathbf{x}_T^{B})$ by maximizing the joint probability of emissions and transitions:
\begin{align}
\mathbf{x}^{B} = \arg \max_{\mathbf{x}_t^{B}} \prod_{t=1}^T \left[ P(\mathbf{x}_t^{B} | \mathbf{x}_t^{A}) \cdot P(\mathbf{x}_t^{B} | \mathbf{x}_{t-1}^{B}) \right],
\end{align}
where $P(\mathbf{x}_t^{B} | \mathbf{x}_t^{A})$ is the emission probability from predictions by $f_{\text{mapping}}$, and $P(\mathbf{x}_t^{B} | \mathbf{x}_{t-1}^{B})$ is the transition probability between target entities (predefined or learned).

\textbf{N-gram} The n-gram model predicts the most probable sequence of target entities $\mathbf{x}^{B} = (\mathbf{x}_1^{B}, \mathbf{x}_2^{B}, \ldots, \mathbf{x}_T^{B})$ by modeling the conditional probability of each target entity given both the corresponding source entity $\mathbf{x}_t^{A}$ and the previous $n-1$ target entities:
\begin{align}
\mathbf{x}^{B} = \arg \max_{\mathbf{x}_t^{B}} \prod_{t=1}^T P(\mathbf{x}_t^{B} | \mathbf{x}_t^{A}, \mathbf{x}_{t-n+1}^{B}, \ldots, \mathbf{x}_{t-1}^{B}),
\end{align}
where $P(\mathbf{x}_t^{B} | \mathbf{x}_t^{A}, \mathbf{x}_{t-n+1}^{B}, \ldots, \mathbf{x}_{t-1}^{B})$ represents the probability of the current target entity $\mathbf{x}_t^{B}$, conditioned on the corresponding source entity $\mathbf{x}_t^{A}$ and the preceding $n-1$ entities in the target sequence ($n$ is predefined as for the naming n-gram as a context length).

\textbf{Decoder-only Transformers} The decoder-only Transformers are also commonly referred to as architectures that are generative pre-trained Transformer (GPT)-like. These models predict the sequence $\mathbf{x}^{B} = (\mathbf{x}_1^{B}, \mathbf{x}_2^{B}, \ldots, \mathbf{x}_T^{B})$ by iteratively generating each token $\mathbf{x}_t^{B}$ based on all previously generated tokens $\mathbf{x}_{1:t-1}^{B}$:
\begin{align}
\mathbf{x}^{B} = \arg \max_{\mathbf{x}^{B}} \prod_{t=1}^{T} P(\mathbf{x}_t^{B} | \mathbf{x}_{1:t-1}^{B}),
\end{align}
where $P(\mathbf{x}_t^{B} | \mathbf{x}_{1:t-1}^{B})$ represents the probability of the \( t \)-th token in the target modality conditioned on its previous tokens, modeled by masked self-attention, which ensures each token in the sequence can only attend to itself and its preceding tokens.

Comparing Viterbi, n-gram, and decoder-only Transformers, the size of the context used for predicting the next token increases progressively (1 for Viterbi, n for n-gram, and Transformers can go very long to thousands of tokens).

Still, these methods have a major drawback in that sequential modeling is only considered for the target modality, which is better addressed by end-to-end sequence-to-sequence neural network models.

\subsubsection{Sequence-to-Sequence Cross-Modality Modeling}

In this type of modeling, the source and target modalities are jointly considered, and the cross-modality mapping function is integrated with sequential modeling into a unified framework. Sequence-to-sequence neural network models \cite{Sutskever2014} ensure that both temporal coherence and inter-modality alignment are achieved.

\textbf{Recurrent Neural Networks (RNNs)} RNNs and their variants, including gated recurrent units (GRUs), and long short-term memory networks (LSTMs), are commonly used for sequence-to-sequence modeling. For cross-modality translation, the process involves encoding the source modality sequence into a latent representation and then decoding it into the target modality sequence.
\begin{itemize}
\item Encoder: The encoder processes the source modality sequence $(\mathbf{x}_{1}^{A}, \mathbf{x}_{2}^{A}, \ldots, \mathbf{x}_{T}^{A})$ to generate a sequence of hidden states $(\mathbf{h}_{1}^{A}, \mathbf{h}_{2}^{A}, \ldots, \mathbf{h}_{T}^{A})$:
\begin{align}
\mathbf{h}_t^{A} = f_{\text{RNN}}(\mathbf{x}_t^{A}, \mathbf{h}_{t-1}^{A}),
\end{align}
where $f_{\text{RNN}}(\cdot, \cdot)$ is the recurrent cell function (e.g., slight variations for vanilla RNNs and GRUs/LSTMs).
\item Decoder: The decoder generates the target modality sequence $(\mathbf{x}_{1}^{B}, \mathbf{x}_{2}^{B}, \ldots, \mathbf{x}_{T}^{B})$ from the encoder's hidden states. Note that the final hidden state $\mathbf{h}_{T}^{A}$ will be treated as a context vector and used to initialize the decoder, i.e., $\mathbf{h}_{0}^{B} = \mathbf{h}_{T}^{A}$. At each time step $t$, the decoder hidden state $\mathbf{h}_t^{B}$ is updated as:
\begin{align}
\mathbf{h}_t^{B} = f_{\text{RNN}}(\mathbf{x}_{t-1}^{B}, \mathbf{h}_{t-1}^{B}),
\end{align}
where $\mathbf{x}_{t-1}^{B}$ is the previous target modality prediction (teacher-forced during training).
\end{itemize}

A classifier can then follow the target modality's hidden states (or the final hidden state) for classification or regression tasks. The model is trained by minimizing the cross-entropy loss between the predicted and ground truth class labels. When the input and output sequences have different lengths, connectionist temporal classification (CTC) loss objective \cite{Graves2006} would replace the standard cross-entropy loss to marginalize all possible alignments of the input sequence to the output sequence.

\textbf{Sequence-to-Sequence Transformers} The sequence-to-sequence Transformer predicts the most probable target sequence $\mathbf{x}^{B} = (\mathbf{x}_1^{B}, \mathbf{x}_2^{B}, \ldots, \mathbf{x}_T^{B})$ given the source sequence $\mathbf{x}^{A}$, as follows:
\begin{align}
\mathbf{x}^{B} = \arg \max_{\mathbf{x}^{B}} \prod_{t=1}^{T} P(\mathbf{x}_t^{B} | \mathbf{x}_{1:t-1}^{B}, \mathbf{h}^{A}),
\end{align}
where $\mathbf{h}^{A}$ represents the contextualized representation of the source sequence modeled by self-attention and cross-attention, $P(\mathbf{x}_t^{B} | \mathbf{x}_{1:t-1}^{B}, \mathbf{h}^{A})$ is the conditional probability of the $t$-th token in the target sequence, predicted by the decoder.

These approaches merge cross-modality mapping and sequential modeling, ensuring both temporal coherence and inter-modality alignment.

\subsection{Multimodal Transformers} \label{sect:transformers}

Transformer architectures have emerged as the dominant neural network framework, offering outstanding performance across diverse domains. Their advantages are rooted in three main characteristics:
\begin{itemize}
\item \textit{Input Tokenization} Transformers are suitable for modeling discrete sequence data of text as a notable advantage over RNNs and their variants, which suffer from vanishing gradient issues, limited memory capacity, and incapability of parallel computation across time steps \cite{Vaswani2017}.
\item \textit{Attention Mechanism} Transformers intrinsically have a more general and flexible modeling space as a notable advantage over CNNs, which are restricted in the aligned grid spaces/matrics \cite{Xu2023}.
\item \textit{Parallel Computing} The inherent parallelism of Transformer computation, driven by matrix operations, ensures scalability for large-scale datasets with faster training speeds.
\end{itemize}

Multimodal Transformers \cite{Xu2023} extend the vanilla architecture to process and fuse information from multiple modalities. We summarize the core mechanisms that make them capable and effective for taking inputs from multiple modalities for fusion tasks. Note that such architectures are different from the decoder-only or the sequence-to-sequence variations described in the previous subsection.

\textbf{Token Fusion} Token fusion approaches include token summation and token concatenation, integrating features from different modalities at the input stage. Token summation considers modality-specific token representations that are summed element-wise, maintaining computational simplicity but limiting modality-specific distinctions. Token concatenation, on the other hand, combines token embeddings into a longer sequence, preserving modality-specific information but increasing computational complexity.

\textbf{Hierarchical Fusion} Hierarchical fusion utilizes attention mechanisms to refine the fusion process by structuring interactions across multiple layers. Multi-stream-to-one-stream hierarchy processes each modality independently before integrating them. It is also viable to adopt a one-stream-to-multi-stream hierarchy that jointly processes inputs and then separates them for fine-tuned representations.

\textbf{Cross-Attention Fusion} Cross-attention extends the vanilla self-attention mechanism by modeling dependencies between tokens from different modalities. Queries from one modality attend to keys and values from another, enabling targeted information exchange. Cross-attention enhances inter-modality interactions while preserving modality-specific features. It is also viable to concatenate two streams of cross-attention and process them by another Transformer to model the global context.

\begin{figure*}[htpb]
\centering
\subfigure[]{\includegraphics[width=.29\linewidth,clip]{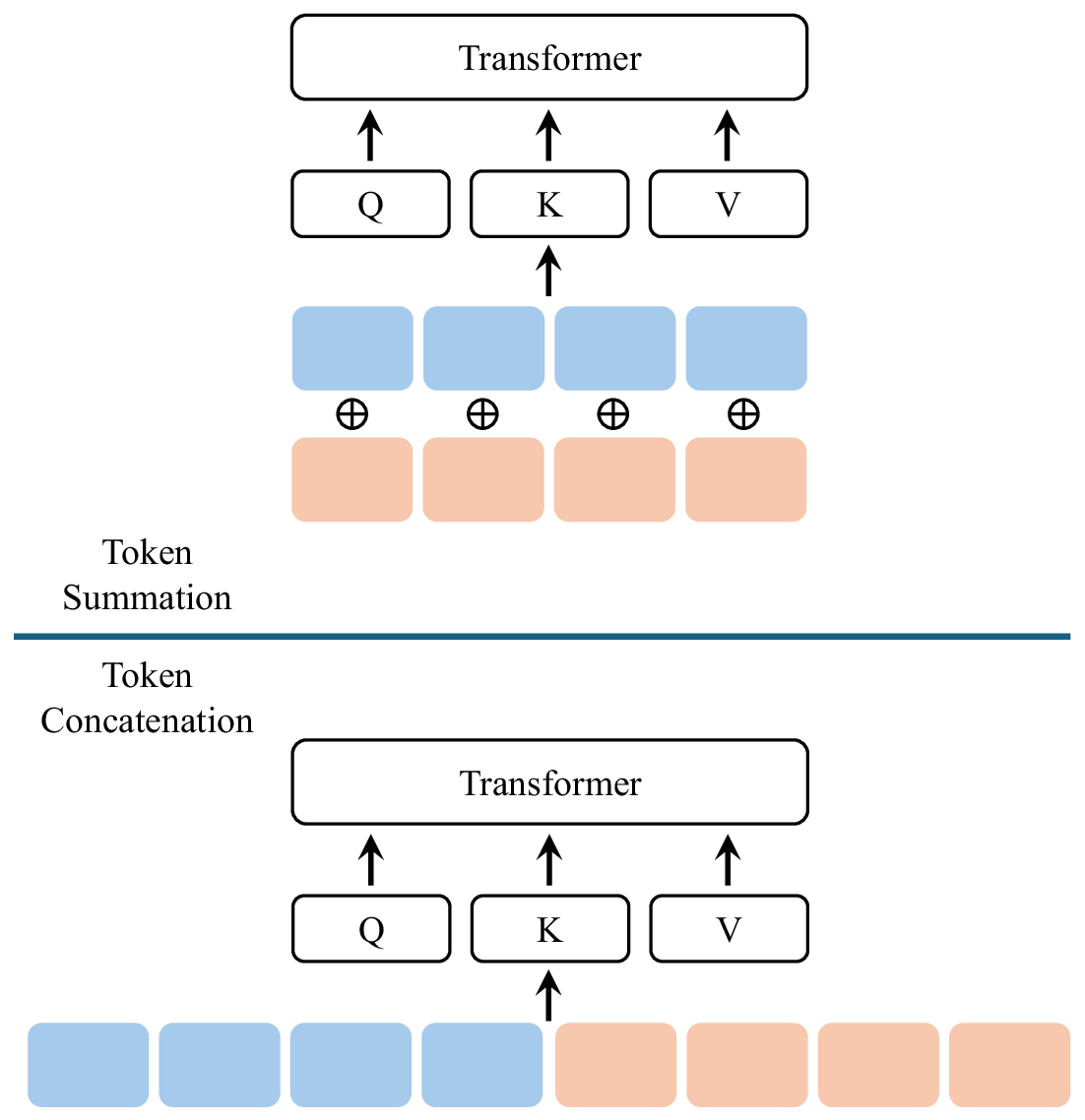}\label{fig:token}}
\subfigure[]{\includegraphics[width=.32\linewidth,clip]{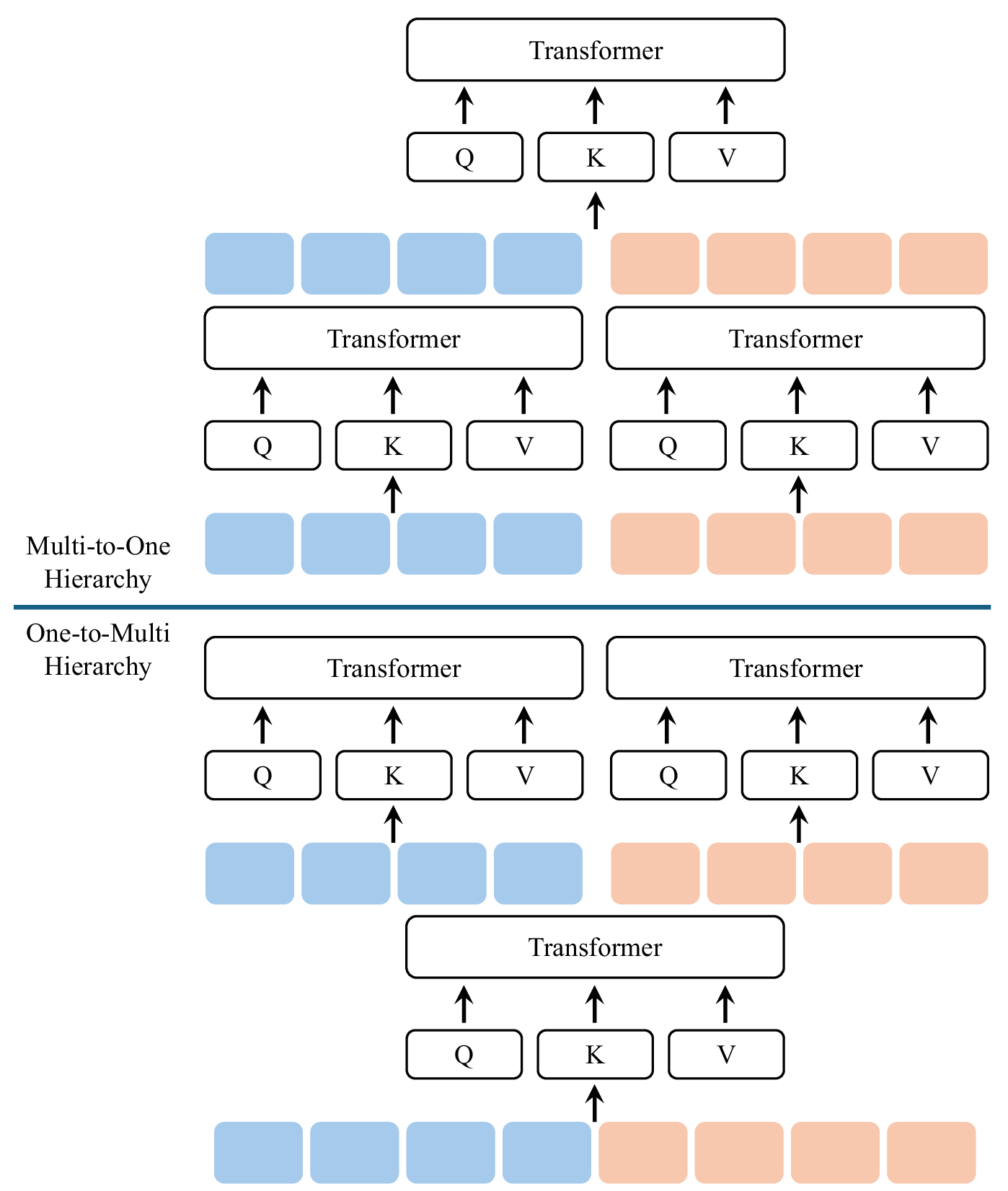}\label{fig:hierarchy}}
\subfigure[]{\includegraphics[width=.32\linewidth,clip]{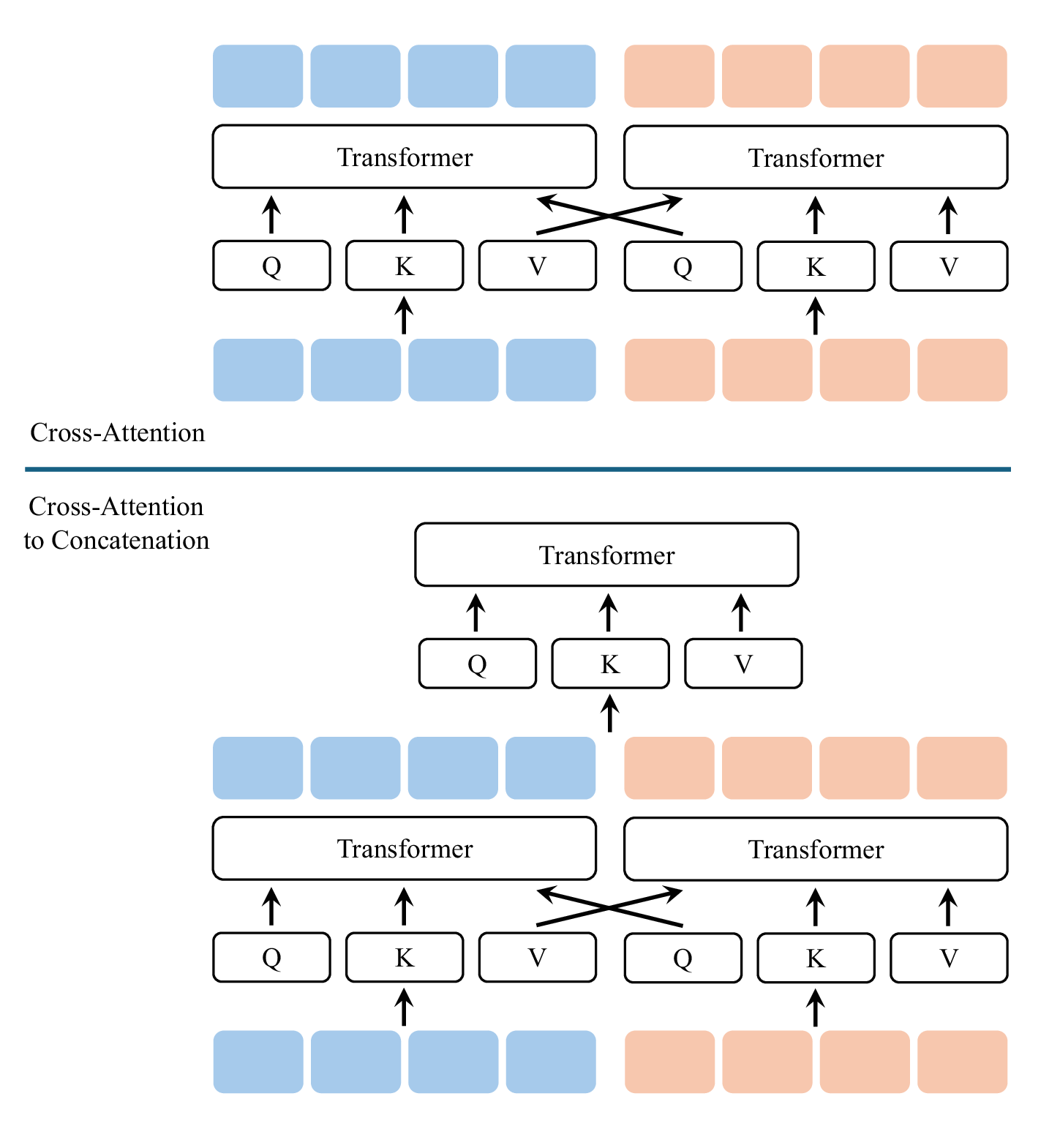}\label{fig:attention}}
\caption{Variants of multimodal Transformers mechanisms that enable multimodal fusion. (a) Token summation and token concatenation; (b) Multi-stream-to-one-stream hierarchy and one-stream-to-multi-stream hierarchy; and (c) Cross-attention and cross-attention to concatenation. Q, K, and V stand for query, key, and value of attention mechanism. Blue and red blocks stand for tokens for two input modalities.} \label{fig:multtransformer}
\end{figure*}

For clarity and to accommodate the wide variations in practical implementations, Fig.~\ref{fig:multtransformer} provides visual illustrations of such concepts mimicking that of the survey by Xu \emph{et al.} \cite{Xu2023}, with exact mathematical formulations omitted for simplicity.

\section{Applications and Datasets} \label{sect:applications}

This section reviews the current literature on representative multimodal BCI applications of visual, speech, and affective decoding.

\subsection{Visual Decoding} \label{sect:vision}

The human visual system has been extensively studied using neurophysiological measurements \cite{Thorpe1996, Kamitani2005}. Early BCI research focused primarily on decoding visual stimuli using paradigms such as P300 potentials \cite{Sutton1965} and steady-state visually evoked potentials \cite{Friman2007}.

This review, however, emphasizes the direct estimation of viewed object categories from recorded brain activity, moving beyond articulately designed visual stimuli. Although this has long been an aspirational objective \cite{Haxby2001}, notable progress has only recently been achieved due to advancements in AI. Recent reviews \cite{Zhang2022, Wilson2024} provide a brief overview and feasibility analysis of this field of visual decoding. In contrast, this subsection delivers an analysis from the algorithmic perspective, covering publicly available datasets, core decoding algorithms, and tasks of interest.

The statistics of publicly available visual decoding datasets are summarized in Table~\ref{tab:data-visual}. The datasets reviewed include those containing public brain data (e.g., EEG or fMRI recorded while subjects viewed images). Whether the input elicitation materials (e.g., ImageNet images or video clips) were publicly available was not considered a criterion for dataset inclusion.

\begin{table*}[htpb]  \center
\caption{Statistics of public visual decoding datasets.}  \label{tab:data-visual}
\scalebox{0.9}{\begin{tabular}{c|c|c|c|c|c|c|c|c|c}
\toprule
\multirow{2}{*}{Dataset} & Brain & Stimulation & \multirow{2}{*}{\# Subjects} & \# Sensors / & Stimulus & \# Total & \# Stimulation & \multirow{2}{*}{\# Classes} & Download \\
 & Data Type & Data Type & & Electrodes & Length & Trials & Materials & & Link \\
\midrule
Kaneshiro2015 \cite{Kaneshiro2015} & EEG & Image & 10 & 124 & 0.5s & 43,206 & 72 & 72 / 6 & \href{https://purl.stanford.edu/bq914sc3730}{SDR} \\
GOD \cite{Horikawa2017} & fMRI & Image & 5 & - & - & 14,750 & 1,250 & 200 & \href{https://figshare.com/articles/dataset/Generic_Object_Decoding/7387130}{FigShare} \\
Video-fMRI \cite{Wen2018} & fMRI & Video & 3 & - & 8mins & - & 972 & 15 & \href{https://purr.purdue.edu/publications/2809/1l}{PU} \\
Shen2019 \cite{Shen2019} & fMRI & Image & 3 & - & - & 21,600 & 1,250 & 200 & \href{https://openneuro.org/datasets/ds001506/versions/1.3.1}{OpenNeuro} \\
BOLD5000 \cite{Chang2019} & fMRI & Image & 4 & - & 1s & 5,254 & 4,916 & 250 / 2000 / 1916 & \href{https://kilthub.figshare.com/articles/BOLD5000/6459449}{FigShare} \\
Grootswagers2019 \cite{Grootswagers2019} & EEG & Image & 16 & 63 & 0.2s & 256,000 & 200 & 200 / 50 / 10 / 2 & \href{https://osf.io/a7knv/}{OSF} \\
NSD \cite{Allen2022} & fMRI & Image & 8 & - & 1s & 213,000 & 10,000 & - & \href{https://naturalscenesdataset.org/}{NSD} \\
THINGS-EEG \cite{Grootswagers2022} & EEG & Image & 50 & 64 & 0.05s & 1,312,400 & 22,248 & 1,854 / 27 & \href{https://osf.io/hd6zk/}{OSF} \\
ThingsEEG \cite{Gifford2022} & EEG & Image & 10 & 64 & 0.1s & 821,600 & 16,740 & 1,854 / 27 & \href{http://www.doi.org/10.17605/OSF.IO/3JK45}{OSF} \\
THINGS-data-fMRI \cite{Hebart2023} & fMRI & Image & 3 & - & 4.5s & 26,220 & 8740 & 720 & \href{https://openneuro.org/datasets/ds004192/versions/1.0.5}{OpenNeuro} \\
THINGS-data-MEG \cite{Hebart2023} & MEG & Image & 4 & 272 & 1.5$\pm$0.2s & 89,792 & 22,448 & 1,854 / 27 & \href{https://openneuro.org/datasets/ds004212/versions/2.0.0}{OpenNeuro}  \\
SEED-DV \cite{Liu2024} & EEG & Video & 20 & 62 & 5mins & - & 1,400 & 40 / 9 & \href{https://bcmi.sjtu.edu.cn/home/eeg2video/}{SJTU} \\
fMRI-Shape \cite{Gao2024} & fMRI & 3D Object & 14 & - & 8s & - & 1,624 & 55 & \href{https://huggingface.co/datasets/Fudan-fMRI/fMRI-Shape}{HuggingFace} \\
fMRI-Objaverse \cite{Gao2024} & fMRI & 3D Object & 5 & - & 6.4s & - & 3,142 & 117 & \href{https://huggingface.co/datasets/Fudan-fMRI/fMRI-Objaverse}{HuggingFace} \\
\bottomrule
\end{tabular}}
\end{table*}

\textbf{Classification Objective} Early studies approached visual decoding as a classification task, training classifiers with brain signals as input and image classes as labels.

Haxby \emph{et al.} \cite{Haxby2001} extracted voxel activity patterns from fMRI for viewing 8 object classes. They used within-category correlations in comparison to between-category correlations for classification. Using 10 object classes, Cox and Savoy \cite{Cox2003} proposed two feature selection approaches for fMRI, selecting voxels with minimum amount of variations across classes, or identifying voxels by comparing responses to whole objects and scrambled images. Cichy \emph{et al.} \cite{Cichy2014} studied both MEG and fMRI responses to 92 object classes. Voxel values from the region of interest were extracted from fMRI as features. For MEG, each time point of the 306-channel MEG was used as feature and analyzed separately. They proposed a binary pairwise classification approach, where the classifier only learns to determine from two available options each time, to differentiate between the real stimuli class from one of the remaining classes.

End-to-end neural networks can be used to directly classify raw EEG signals. Bagchi \emph{et al.} \cite{Bagchi2022} proposed a CNN+Transformer network. They use multi-headed self-attention and temporal convolution to enhance architecture over classic CNN architectures. Similarly, Luo \emph{et al.} \cite{Luo2024a} proposed a dual-branch spatio-temporal-spectral Transformer feature fusion network. They utilized channel attention and graph convolutions for better performance.

Besides explicit classification for objects, human perception can also be analyzed for decoding other kinds of mental states. Haynes and Rees \cite{Haynes2005} studied orientations, using only two object classes as a stimulus of contrast-reversing gratings with orthogonal orientation. With simple voxel values under feature selection and classic classifiers, they showed that fMRI-based decoding can obtain a direct measure of orientation-selective processing. Kamitani and Tong \cite{Kamitani2005} studied 8 stimulus orientations with a similar approach.

\textbf{Retrieval Objective} Cross-modality learning transforms classification tasks into retrieval objectives, aligning representations across modalities. This would require using representations of the stimulus materials, which can usually be from models pre-trained on large-scale image datasets or image-text datasets such as CLIP (which stands for contrastive language-image pre-training) \cite{Radford2021} model.

Horikawa and Kamitani \cite{Horikawa2017} proposed to use a linear regression function that transformed fMRI patterns into CNN features of the viewed images. The novel image categories can be estimated by comparing the correlation coefficients between the predicted image features and the category-average visual features of various categories. Chen \emph{et al.} \cite{Chen2024a} proposed to cast the mapping problem as an optimal transport problem. They introduced a graph matching framework for the alignment of fMRI and image representations. Song \emph{et al.} \cite{Song2024} adopted the InfoNCE objective, and further utilized self-attention and graph attention to enhance existing architectures using EEG or MEG. Gong \emph{et al.} \cite{Gong2024} extracted voxel features based on discrete Fourier transform, and further aligns with image representations with InfoNCE objective and diffusion model. Zhou \emph{et al.} \cite{Zhou2024} introduced representational similarity analysis for modality alignment. Representations of blood-oxygen-level dependent of fMRI across subjects were extracted by Transformer model, to align with image representation in the space of pretrained image model \cite{Radford2021}.

\textbf{Generative Mapping in a Joint Latent Space} Du \emph{et al.} \cite{Du2017} proposed a deep generative multiview model for joint latent space modeling. The reconstruction of the perceived image was cast as the problem of Bayesian inference of the missing modality. Each modality can also be modeled through different types of models, where neural networks can be used for images, and a sparse Bayesian linear generative model for fMRI \cite{Du2019}. Han \emph{et al.} \cite{Han2019} utilized a VAE for joint latent space construction. The VAE maps the image to the latent space, whereas a linear regression model maps fMRI responses to the VAE’s latent variables, aligning the two modalities into a unified representation for decoding and reconstruction. Cheng \emph{et al.} \cite{Cheng2023} used a generator in the form of GAN or diffusion model for joint latent space alignment. Beyond natural images, they also studied the stimulus of illusory lines and neon color spreading to understand the subjective experience of visual perception. Qian \emph{et al.} \cite{Qian2024} used ridge regression to constrain the learning space for joint latent space modeling. To learn a better global representation of fMRI, they also used masked autoencoder \cite{He2022} that is able to reconstruct from only the class token.

\textbf{Generative Mapping in a Conditional Latent Space} Zeng \emph{et al.} \cite{Zeng2024} utilized diffusion model for conditional latent space construction. The fMRI signals act as conditioning inputs to the latent diffusion model and directly guide the image generation process in terms of both semantics and spatial structure. The latent diffusion model generates images by explicitly conditioning the denoising process on the extracted fMRI features. Qian \emph{et al.} \cite{Qian2024a} proposed to use the diffusion model with CNN+Transformer architecture for conditional latent space modeling. Scotti \emph{et al.} \cite{Scotti2023} used InfoNCE objective with mixup data augmentation \cite{Zhang2018Mixup} for image reconstruction with fMRI. The objective was then further softened with knowledge distillation and regularized with diffusion prior. Benchetrit \emph{et al.} \cite{Benchetrit2024} used the InfoNCE objective to align MEG and image representations. They also used the diffusion model to convert the latent representation back to the stimulus image object class. Li \emph{et al.} \cite{Li2024} also proposed to directly reconstruct viewed image from brain data, using InfoNCE loss and diffusion model, verified for EEG, MEG, and fMRI, respectively. Chen \emph{et al.} \cite{Chen2023} used an autoencoder with masked signal modeling as a pretext task to learn fMRI representations. The trained fMRI encoder is then integrated with a latent diffusion model through cross-attention and time-step conditioning for conditional latent space modeling.

\textbf{Text Caption-aided Visual Decoding} Text captions can be generated for the images of stimulus materials, using pre-trained vision-language models \cite{Zhang2024VLM}. Being an alternative option other than images, such texts' representations can further help establish a better cross-modality mapping function.

The integration of text captions into visual decoding frameworks leverages the powerful capabilities of pre-trained vision-language models \cite{Zhang2024VLM} to enrich the cross-modal mapping process. By incorporating textual descriptions associated with visual stimuli, these approaches provide an additional modality that bridges the gap between brain signals and visual content, enabling more robust and interpretable decoding.

Du \emph{et al.} \cite{Du2023} proposed a VAE-based model that works in two collaborative parts, i.e., multi-modality joint modeling and mutual information regularization, to learn the relationship between brain activities, visual stimuli, and Wikipedia descriptions of classes. Then, a support vector machine is used to perform the classification task with the learned latent representation. Similarly, Liu \emph{et al.} \cite{Liu2023} also adopted the InfoNCE objective for fMRI-to-image or fMRI-to-text retrieval. Xie \emph{et al.} \cite{Xie2024} projected the fMRI data into image and text latent spaces using a diffusion prior, maintaining brain-visual-linguistic consistency. Xia \emph{et al.} \cite{Xia2024} utilized InfoNCE objective with mixup data augmentation to learn fMRI-image-text shared representations.

Takagi and Nishimoto \cite{Takagi2023} used linear regression to map brain activity to image and text latent spaces, with early visual cortex signals providing low-level structural details and higher visual cortex signals offering semantic context. The latent diffusion model then refined noisy latent into coherent images, conditioned on both structural and semantic features, enabling conditional latent space generative reconstruction.

Not specifically aiming for decoding the visual stimulus, Wang \emph{et al.} \cite{Wang2023BrainBERT} investigated the ability of multimodal representations from pre trained vision-language models to predict high-level visual cortex responses in the human brain. They found that the inclusion of text feedback alongside diverse and large-scale datasets allows better comprehension of brain data, which aligns well with neural representations in high-level brain areas.

\textbf{Beyond Image Decoding} Decoding visual experiences from brain data extends beyond static images to include dynamic stimuli such as videos and 3D objects. This extension broadens the scope of applications, allowing researchers to decode and reconstruct richer visual experiences that align more closely with real-world perception.

Li \emph{et al.} \cite{Li2024c} proposed reconstructing video stimulus from fMRI data, aligning functional and anatomical brain activity into a unified space. In contrast to traditional fMRI decoding methods that flatten each frame of fMRI into a 1D signal and select subject-specific activated voxels, their approach transformed fMRI into a unified representation across subjects into the latent space of the pre-trained vision-language model. The fMRI embedding was utilized as input to the cross-attention module of the diffusion model for conditional latent space modeling.

Lu \emph{et al.} \cite{Lu2024} also proposed reconstructing video stimulus from fMRI data. Semantic, structural, and motion information features are projected into the latent space using InfoNCE objective, vector quantized VAE, and a Transformer model, respectively. A latent diffusion model then reconstructs video stimulus from such three features from the conditional latent space. Liu et al. \cite{Liu2024} proposed to reconstruct dynamic videos from EEG signals. Utilizing the much higher temporal resolution of EEG signals (compared to fMRI), a sequence-to-sequence Transformer model aligned the EEG and video representations. Text captions and their representations were generated for video frames using a pre-trained vision-language model \cite{Li2022BLIP, Radford2021}, and then aligned with EEG representations. Reconstruction was then refined with a diffusion model to ensure consistency across generated video frames.

Gao \emph{et al.} \cite{Gao2024, Gao2024a} proposed to reconstruct 3D objects as stimuli from fMRI. They proposed to align the representations of fMRI and video frames of 3D objects in the latent space of pre-trained vision-language model \cite{Radford2021}. The fMRI feature was then used to condition the diffusion model. Finally, a vector quantized decoder converts the latent vector back into 3D mesh representation. Guo \emph{et al.} \cite{Guo2024} considered 3D object reconstruction from EEG of 72 classes to recover both the shape and color of 3D objects. They utilized InfoNCE objective, dynamic-static EEG-fusion encoder, and a colored point cloud decoder.

\textbf{Application Value} Despite the significant progress in decoding visual stimuli from brain data, the real-world applications of such technologies remain relatively under-explored. Unlike paradigms like P300 or SSVEP, which are well-established for use in spellers or other practical BCIs, visual decoding primarily serves as a tool for understanding neuromechanisms rather than addressing immediate functional needs.

One potential avenue lies in developing assistive technologies for individuals with visual or cognitive impairments. By decoding their internal visual experiences, these systems could facilitate communication or enhance accessibility \cite{Beauchamp2020}. Another promising direction is in immersive entertainment and virtual reality, where brain-driven visual decoding could provide personalized and adaptive experiences \cite{Ferrante2024}.

However, challenges remain in bridging the gap between research and application. The high computational demands, reliance on specialized equipment, and lack of accuracy limit the widespread adoption of visual decoding technologies. Continued explorations in decoding algorithms and practical applications are crucial to unlocking the full potential of visual decoding-based BCIs.

\subsection{Speech Decoding} \label{sect:language}

Language is the primary medium through which humans communicate complex ideas, emotions, and intentions. Understanding how the brain encodes and processes language is a central question in neuroscience, with profound implications for BCIs. A prominent application is speech neuroprostheses \cite{Chang2024}, which aims to restore communication by decoding neural activity into intelligible speech or text. Unlike other BCI tasks, speech decoding must address the sequential and dynamic nature of linguistic information, where components like phonemes and words unfold over time.

A recent review by Silva \emph{et al.} \cite{Silva2024} provides a comprehensive overview of speech neuroprostheses, primarily on the used features and practices, while this review focuses on the AI algorithms that enable sequential cross-modality translation. Sequential modeling approaches, which are only briefly discussed in Mathis \emph{et al.} \cite{Mathis2024}, are examined. This review also extends beyond speech neuroprostheses to general linguistic tasks, such as reading and listening. Table~\ref{tab:data-speech} summarizes the statistics of publicly available speech decoding datasets.

\begin{table*}[htpb]
\centering
\caption{Statistics of public speech decoding datasets.}  \label{tab:data-speech}
\scalebox{0.88}{\begin{tabular}{c|c|c|c|c|c|c|c|c}
\toprule
\multirow{2}{*}{Dataset} & Brain Data & \multirow{2}{*}{\# Subjects} & \# Sensors / & \multirow{2}{*}{Language} & \multirow{2}{*}{Duration} & \multirow{2}{*}{Task Type} & \multirow{2}{*}{Disease} & Download /  \\
 & Type & & Electrodes & & & & & Request Link \\
\midrule
ZuCo \cite{Hollenstein2018} & EEG & 12 & 128 & English & 60h & Reading & None & \href{https://osf.io/q3zws/}{OSF} \\
Broderick2018 \cite{Broderick2018} & EEG & 19 & 128 & English & 19h & Listening & None & \href{https://doi.org/10.5061/dryad.070jc}{Dryad} \\
Brennan-Hale2019 \cite{Brennan2019} & EEG & 33 & 60 & English & 6.8h & Listening & None & \href{https://deepblue.lib.umich.edu/data/concern/data_sets/bn999738r}{UMich} \\
MOUS-MEG \cite{Schoffelen2019} & MEG & 204 & 275 & Dutch & - & Reading / Listening & None & \href{https://data.ru.nl/collections/di/dccn/DSC_3011020.09_236}{RU} \\
MOUS-fMRI \cite{Schoffelen2019} & fMRI & 204 & - & Dutch & - & Reading / Listening & None & \href{https://data.ru.nl/collections/di/dccn/DSC_3011020.09_236}{RU} \\
ZuCo 2.0 \cite{Hollenstein2020} & EEG & 18 & 105 & English & 42h & Reading & None & \href{https://osf.io/2urht/}{OSF} \\
Verwoert2022 \cite{Verwoert2022} & SEEG & 10 & 54-127 & Dutch & 33min & Reading & Epilepsy & \href{https://osf.io/nrgx6/}{OSF} \\
MEG-MASC \cite{Gwilliams2023} & MEG & 27 & 208 & English & 56.2h & Listening & None & \href{https://osf.io/ag3kj/}{OSF} \\
Willett2023 \cite{Willett2023} & MEA & 1 & 128 & English & 54h & Attempted Speech & Amyotrophic Lateral Sclerosis & \href{https://doi.org/10.5061/dryad.x69p8czpq}{Dryad} \\
Metzger2023 \cite{Metzger2023} & ECoG & 1 & 253 & English & 22.4h & Attempted Speech & Anarthria \& Quadriplegia & \href{https://doi.org/10.5281/zenodo.8200782}{Zenodo} \\
Chisco \cite{Zhang2024b} & EEG & 3 & 125 & Mandarin Chinese & 45h & Imagined Speech & None & \href{https://openneuro.org/datasets/ds005170}{OpenNeuro} \\
ChineseEEG \cite{Mou2024} & EEG & 10 & 128 & Mandarin Chinese & 110h & Silently Reading & None & \href{https://openneuro.org/datasets/ds004952}{OpenNeuro} \\
Du-IN \cite{Zheng2024} & SEEG & 12 & 72-158 & Mandarin Chinese & 36h & Reading & Epilepsy & \href{https://huggingface.co/datasets/liulab-repository/Du-IN}{HuggingFace} \\
\bottomrule
\end{tabular}}
\end{table*}

\textbf{Instance-wise Classification} Preliminary research on speech decoding primarily established the feasibility of mapping brain data to language elements, either by demonstrating correlations \cite{Bouchard2013, Chartier2018, Dichter2018, Leonard2024} or by employing rudimentary classification methods for mapping brain recordings to phonemes and words.

Angrick \emph{et al.} \cite{Angrick2024a} studied speech onset detection using ECoG from an ALS participant with dysarthria under attempted speech task in English. A binary LSTM model was used for direct binary classification. Mugler \emph{et al.} \cite{Mugler2014} used ECoG and corresponding time-frequency features with classic classifiers for decoding into phonemes under reading task in English. Lee \emph{et al.} \cite{Lee2020} used EEG for decoding into 12-word vocabulary under imaged speech task in English. Common spatial pattern feature and classic classifier was used. Luo \emph{et al.} \cite{Luo2023} decoded ECoG of an ALS patient under reading and miming task in English. CNN was used to classify ECoG into 6 words of directional commands. Wandelt \emph{et al.} \cite{Wandelt2024} decoded MEA under internal speech and reading task for six-word and two-pseudoword vocabulary in English. Neural firing rates were used as features under classic classifiers. Tan \emph{et al.} \cite{Tan2024} proposed a hyperbolic neural network for direct classification into phonemes using MEA recorded with 96-channel Utah arrays under Mandarin Chinese reading task. They used a hierarchical clustering loss objective to guide the representation learning. Liu \emph{et al.} \cite{Liu2023a} used ECoG to classify 8 tonal syllables under Mandarin Chinese reading task. CNNs were used to decode lexical tones and base syllables independently.

\textbf{Instance-wise Retrieval} Recent advances leverage cross-modality contrastive learning and generative modeling to expand the label space and accommodate novel classes in retrieval task. These approaches use instance-wise mapping functions to align neural and linguistic representations.

D{\'e}fossez \emph{et al.} \cite{Defossez2023} utilized cross-modality contrastive learning using MEG and EEG data under listening task in English or Dutch. Under the InfoNCE objective, neural activity representations were aligned with deep speech representations, obtained from a pre-trained speech model. Duan \emph{et al.} \cite{Duan2023} integrated discrete encoding sequences into open-vocabulary EEG-to-text translation with pre-trained language models under reading task in English. Using both word-level features and raw EEG signals, they applied a combination of the InfoNCE objective and self-reconstruction loss objective to improve alignment between EEG and text representations.

Zheng \emph{et al.} \cite{Zheng2024} proposed a discrete codex-guided mask modeling framework for instance-wise decoding 61 words from SEEG under reading task in Mandarin Chinese. Using a conditional latent space, the model employs Transformer model under temporal modeling and mask modeling tasks for pre-training. Ma \emph{et al.} \cite{Ma2025} proposed a multi-scale cycle-consistent dual generative adversarial network for EEG/SEEG-to-speech translation under listening task in English. For cross-modality mapping in a conditional latent space, their model employed a cycle-consistency loss objective, aligning the temporal and spectral representations of EEG/SEEG signals and speech waveforms.

\textbf{Target-Centric Sequential Modeling via Classic Language Models} As a fundamental element of language, coherence across phonemes and words is the core requirement to form concrete communication. Sequential modeling is thus a must-have for better-performing speech decoding, considering contextual dependencies and allowing the decoded output to align more closely with natural language patterns.

Herff \emph{et al.} \cite{Herff2015} used Viterbi algorithm on language models to improve the coherence of sequences of phonemes, using ECoG data under reading task in English. Their model showed high error rates under increased vocabulary sizes. Sun \emph{et al.} \cite{Sun2020} proposed a character-level decoding approach from ECoG under reading task in English. Their model took ECoG signals as input, using 3D inception layers for multiband spatiotemporal feature extraction, dilated CNNs (a method to extend CNNs for sequential correlations like RNNs) for sequence learning, and a CTC objective optimized for character decoding. For sequential modeling, they incorporated a language model in the form of either n-gram or RNN to evaluate the likelihood of specific character sequences with an over 1,000-word vocabulary. Moses \emph{et al.} \cite{Moses2021} used long-term ECoG of a paralyzed individual with anarthria to decode words and sentences under attempted speech task of 50-word vocabulary in English. They used an LSTM-based speech detection model to identify speech onset, a bidirectional GRU-based word classification model to classify 50 words, and an n-gram language model combined with a Viterbi decoder for sentence reconstruction. Zhang \emph{et al.} \cite{Zhang2024c} proposed a modular neural network framework using ECoG under reading task in Mandarin Chinese. The framework combined three components of speech onset detection, tone (four tones in Mandarin Chinese) decoding, and syllable decoding modules, integrated post-hoc using a Viterbi-based Bayesian language model for sequential modeling.

\textbf{Language Modeling via Large Language Models} Large Language Models (LLMs) have become the standard in language modeling due to its unprecedented performance in all language tasks \cite{Brown2020, Zhao2023LLM}. The decoder-only GPT-like LLMs thus replace n-gram or Viterbi in language-wise sequential modeling. Additionally, they are also capable for handling much more complicated cases such as multi-lingual decoding.

Mischler \emph{et al.} \cite{Mischler2024} used intracranial EEG data collected under listening task in English. They applied ridge regression to align representations of neural responses to corresponding text data, obtained with open-source LLMs' representations. They demonstrated that higher-performance LLMs align more closely with brain-like hierarchical feature extraction pathways, particularly benefiting from contextual information.

Tang \emph{et al.} \cite{Tang2023a} proposed to decode fMRI to language under perceived or imagined speech task in English. They used a multivariate Gaussian distribution modeling for conditional latent space cross-modality instance-wise generative modeling. Then, they used a fine-tuned LLM to predict coherent sequences, additionally leveraging beam search to optimize candidate sequences for vocabulary consisting of thousands of words.

Tang \emph{et al.} \cite{Tang2024} used ECoG from stroke patients with dysarthria under silently attempted speech task in English. Brain data were segmented into tokens, where a 1D-CNN directly classified each segment into a corresponding language token. For sequential modeling, a pre-trained LLM then refined sentences by incorporating contextual and emotional cues.

Silva \emph{et al.} \cite{Silva2024a} decoded ECoG data from a participant with anarthria and paralysis into sentences under bilingual attempted speech in English and Spanish. They employed a bidirectional GRU for word classification using ECoG signals. For sequential modeling, they used language-specific n-gram and a bilingual large language model, followed by a bilingual beam search to refine word sequences and flexibly infer the intended language.

\textbf{Direct Sequence-to-Sequence Translation} Sequence-to-sequence modeling has emerged as a powerful framework for speech decoding. It involves transforming a temporal sequence of neural signals into another sequential representation, such as phonemes, words, or sentences, by leveraging encoder-decoder architectures of RNNs/GRUs/LSTMs or encoder-decoder Transformers. The encoder processes raw brain signal data, extracting meaningful high-dimensional representations, while the decoder generates structured output sequences by maintaining temporal coherence and contextual relevance. These models can handle the inherent temporal dynamics of neural activity, accommodating the sequential unfolding of linguistic information over time.

Makin \emph{et al.} \cite{Makin2020} developed an encoder-decoder framework for word-level cross-modality translation using ECoG under reading task in English. High-gamma features of ECoG signals were processed with temporal convolutions, encoded into high-dimensional representations by an RNN. For sequential modeling, another decoder RNN used this representation to generate words of a sentence, maintaining coherence between successive predictions. Duraivel \emph{et al.} \cite{Duraivel2023} decoded ECoG to phoneme sequences under speech repetition tasks in English. A sequence-to-sequence RNN was used to map ECoG features to phoneme sequences, one phoneme at a time, using the encoder's output.

Willett \emph{et al.} \cite{Willett2023} and Card \emph{et al.} \cite{Card2024} both studied MEA for text decoding for a (different) participant with ALS under attempted speech task in English. Willett \emph{et al.} \cite{Willett2023} used a GRU to map spike features to phoneme probabilities, leveraging CTC objective to align neural inputs and phoneme sequences. Decoded phonemes were sequentially modeled with an n-gram language model via Viterbi search to infer fluent word sequences. In Card \emph{et al.} \cite{Card2024}, neural signals were processed to extract spike-band power and threshold crossings. A bidirectional GRU was used under CTC objective in an identical purpose. The outputs were integrated with a n-gram language model, followed by multi-stage rescoring, to construct fluent word sequences. Both systems were evaluated for both 50-word and 125,000-word vocabulary and achieved low error rates.

\textbf{Acoustic Speech Waveform Generation} Direct acoustic speech waveform conversion bypasses intermediate linguistic representations like phonemes or words by translating brain signals to continuous acoustic representations. Mel-spectrograms, which capture the spectral characteristics of speech in a time-frequency domain, are commonly used as intermediary representations. Each temporal segment of neural activity is mapped to and aligned with corresponding frames in the mel-spectrogram. Vocoders, short for voice coders, then convert the mel-spectrogram into raw audio/speech waveforms. Articulatory kinematics and linear predictive coefficients are also frequently employed as intermediate representations. The vocoders are usually in the architecture of sequence-to-sequence RNNs, such as WaveNet \cite{VanDenOord2016} and its variants.

Angrick \emph{et al.} \cite{Angrick2019} used ECoG data for speech synthesis under reading task in English. They employed densely connected 3D CNN \cite{Huang2017} to map spatiotemporal feature of broadband gamma power of ECoG onto logMel spectrograms, which provide a compressed acoustic representation of speech. A pre-trained WaveNet vocoder conditioned on logMel features\cite{Shen2018} then reconstructed speech waveforms.

Anumanchipalli \emph{et al.} \cite{Anumanchipalli2019} used ECoG data for speech synthesis under reading and miming task in English. They utilized a bidirectional LSTM model to decode articulatory kinematics from neural signals, followed by a second LSTM to convert kinematics into acoustic features for synthesizing speech waveform.

Metzger \emph{et al.} \cite{Metzger2023} decoded ECoG of a participant with severe limb and vocal paralysis under attempted speech task in English. Three output modalities: text, synthesized speech, and facial avatar animations, were considered. Signals were mapped into phonemes through a bidirectional GRU, n-gram model and beam search were then used to refine the phoneme predictions for sequential modeling. The phonemes were then converted into discrete speech units from mel-spectrograms of speech-sound features, and the speech waveform was further synthesized using a unit-to-speech pre-trained vocoder \cite{Lee2022a}. Articulatory gestures, which were converted using an encapsulated model for animating live speech, were then used to animate synchronized facial movements of an avatar in real-time.

Angrick \emph{et al.} \cite{Angrick2024} used ECoG from an ALS participant for decoding a closed 6-word vocabulary under attempted speech task in English. The decoding pipeline included three RNNs: a one-directional one to identify speech segments, a bidirectional one to map signals onto linear predictive coefficients, and a pre-trained vocoder \cite{Valin2019} to reconstruct acoustic waveforms.

Wairagkar \emph{et al.} \cite{Wairagkar2024} used MEA from a participant with ALS and dysarthria under attempted speech task in English. Acoustic features were extracted using a Transformer model. These features included vocal tract, tone, and voicing characteristics, which were converted into continuous speech by a pre-trained vocoder \cite{Valin2019}.

Chen \emph{et al.} \cite{Chen2024} used ECoG data from epilepsy participants for speech synthesis under a series of speech tasks in English. ECoG signals were mapped into acoustic speech parameters through an ECoG encoder with 3D ResNet, 3D Swin Transformer \cite{Liu2021}, or LSTM architectures. A differentiable speech synthesizer then converted these parameters into a spectrogram, which was subsequently transformed into speech waveforms.

\textbf{Application Broadness} Speech neuroprosthesis has transformative potential for individuals with severe speech impairments, offering a pathway to regain communication capabilities. These technologies can bridge the gap for patients with conditions like ALS, anarthria, or stroke-induced paralysis, providing them with a means to express themselves and interact with the world. The societal implications of restoring speech extend beyond basic communication, encompassing emotional expression, social engagement, and improved quality of life.

Despite their promise, the applicability of speech neuroprostheses remains limited. Current methods heavily rely on invasive brain-recording techniques, such as ECoG or MEA, which require surgical implantation and long-term use to collect subject-specific training data \cite{Vansteensel2024}. This restricts accessibility to individuals who meet strict medical criteria. Additionally, decoding performance often diminishes for less invasive methods like EEG, due to lower signal resolution and susceptibility to noise. The lack of scalable, non-invasive solutions hinders broader adoption. Additionally, the performance of direct waveform conversion pipelines, particularly those using mel-spectrograms, also remains underwhelming, limiting the naturalness and intelligibility of synthesized speech. Continued research in decoding accuracy, model generalization, and minimum calibration strategies are essential for enabling the widespread adoption of speech decoding-based BCIs.

\subsection{Affect Decoding} \label{sect:emotion}

Affect decoding is becoming increasingly significant due to its essential role in human cognition, communication, and decision-making processes \cite{Lerner2015}, as well as its critical importance for intelligent systems \cite{Minsky1988}. An affective BCI is defined as a system that monitors and/or regulates the emotional states of the brain \cite{Wu2023}. While our prior review comprehensively addressed the purpose, datasets, and machine learning components of affective BCIs, this review focuses specifically on fusion methods that leverage contemporary AI models under multimodal inputs. The publicly available datasets are omitted in this review as they were summarized in \cite{Wu2023}, including information on the specific modalities.

\textbf{Feature- or Decision-Level Fusion} Feature- and decision-level fusion methods represent classic approaches for integrating multiple modalities. Feature-level fusion typically involves concatenating features extracted from different modalities into a unified representation, which is then used for subsequent classifier training. In contrast, decision-level fusion independently trains classifiers for each modality, with predictions aggregated through ensemble strategies based on prediction scores.

For feature-level fusion, Wu \emph{et al.} \cite{Wu2010} concatenated features from four physiological signals (EEG, electrocardiogram, skin conductance level, and respiration) for arousal classification. Zheng and Lu \emph{et al.} \cite{Zheng2017} concatenated EEG and electrooculography features for vigilance regression. Soleymani \emph{et al.} \cite{Soleymani2012} combined EEG signals, eye gaze data, and pupillary responses to classify emotions. Beyond feature-level fusion, they also explored decision-level fusion where classifications from each modality are combined based on their confidence (prediction probability) scores.

Chen \emph{et al.} \cite{Chen2016} introduced a framework for integrating multimodal and multiset neurophysiological data. They used complementary modalities of EEG, fMRI, electromyography, and kinematic data to enhance brain function analysis. They introduced a joint blind source separation approach that achieves multimodal fusion by recovering ensembles of independent sources while exploiting shared statistical dependencies across datasets. This fusion methodology addresses the limitations of single-modality analyses by maximizing the correlation or independence across modalities through advanced optimization techniques like canonical correlation analysis and its variants.

Gong \emph{et al.} \cite{Gong2024a} integrated EEG and eye movement modalities for emotion classification. They used a non-neural network approach called completeness modality representation learning to capture both modality-independent and modality-relevant features. Additionally, they used a weighted representation distribution alignment to align marginal and conditional distributions across subjects.

For a comprehensive survey of these methods, Poria \emph{et al.} provided a detailed review that highlights their applications in affective computing.

\textbf{Representation Fusion} Representation fusion leverages neural networks to integrate feature extraction and fusion processes, enabling the automatic learning of multimodal representations and fusion of multiple modalities simultaneously.

Han \emph{et al.} \cite{Han2015} integrated fMRI-based brain responses and low-level audio-visual features for arousal classification. Multimodal fusion was achieved using a multimodal deep Boltzmann machine, which learns a joint probabilistic representation of the modalities. The model captured the complementary features of fMRI-derived functional connectivity matrices and audio-visual characteristics, enabling effective classification while allowing inference of joint representations from audio-visual features alone during testing.

Du \emph{et al.} \cite{Du2018b} utilized EEG and eye movement signals for emotion classification. The fusion of modalities is achieved through a multi-view variational autoencoder framework that models the statistical relationships between modalities in a joint latent space. A non-uniformly weighted Gaussian mixture posterior approximates the latent space, allowing dynamic weighting of each modality based on its importance, which facilitates robust semi-supervised classification and imputation of missing data. This approach leverages both labeled and unlabeled data to improve performance on incomplete multi-modal datasets with missing modalities.

Zheng \emph{et al.} \cite{Zheng2019} integrated EEG and eye movement data for emotion classification. A bimodal autoencoder extracted shared high-level representations from both modalities. The framework first trained individual restricted Boltzmann machines for EEG and eye movement features, then combined them into a shared representation through stacking and fine-tuning.

Yan \emph{et al.} \cite{Yan2021} used EEG and eye movement data for emotion classification. The fusion was achieved with a bimodal autoencoder for conditioning on a joint latent space. Thus representation of the missing modality could be synthesized using a conditional GAN from single-modality eye movement data, reducing reliance on EEG.

\textbf{Fusion via Multimodal Transformers} Multimodal Transformers have emerged as powerful tools for integrating information across diverse modalities, due to advantages detailed in Section~\ref{sect:transformers}.

Jiang \emph{et al.} \cite{Jiang2023} proposed a multimodal Transformer for emotion classification using EEG and eye movement modalities. They used multi-stream to one-stream hierarchical attention, where EEG and eye movement features are processed independently within their respective streams using self-attention, followed by a fusion mechanism that integrates the outputs of these streams into a unified representation. Adversarial training is further applied to improve cross-subject generalization by reducing domain-specific discrepancies between subjects.

Li \emph{et al.} \cite{Li2024b} integrated acoustic, textual, and EEG modalities for spoken language intent recognition, using intent categories (e.g., query, irony, praise) as the task labels. To model inter-modal dependencies, multi-head cross-attention of multimodal Transformers dynamically adjusted the weight of acoustic-textual joint features using EEG-based brain representations. Additionally, the framework employed a joint multi-task learning approach, optimizing intent recognition as the primary task and emotion recognition as an auxiliary task to enhance performance. This multimodal strategy enabled the interpretation of the speaker’s true intent.

Yin \emph{et al.} \cite{Yin2024} used EEG and electrooculography modalities for emotion classification. The fusion was achieved through multimodal Transformers with hierarchical attention. Intra-modal channel attention focused on extracting dependencies and assigning weights across different EEG and electrooculography channels within each modality, while cross-modal temporal attention captured global dependencies across the temporal features of EEG and electrooculography signals, allowing for the integration of information across modalities.

\textbf{Sequential Fusion} Sequential fusion methods focus on leveraging temporal correlations to enhance multimodal analysis. Such modes attempt to integrate representation learning, sequential modeling, and multimodal fusion in a unified framework. Sequential neural network models and Multimodal Transformers are widely adopted for this purpose.

Zhu \emph{et al.} \cite{Zhu2024} integrated EEG and facial expression data for valence and arousal classification. They proposed a self-attention-based multi-channel LSTM for temporal feature alignment and a confidence regression network to estimate the reliability of each modality. The fusion process dynamically weighted modalities based on their confidence levels and combined them using a weighted concatenation strategy.

Zhang \emph{et al.} \cite{Zhang2024a} utilized blood volume pulse and electrodermal activity signals for stress recognition, with binary and multi-class stress labels. 1D-CNNs and LSTMs were used to model these two modalities, respectively. The fusion was achieved using a cross-attention that aligned and fused the two modalities to learn joint representations under temporal slices, while self-attention reduced noise.

Koorathota \emph{et al.} \cite{Koorathota2022} employed EEG, photoplethysmography, and electrodermal activity data for valence and arousal classification. Multimodal Transformers used cross-attention to align and integrate features across modalities, while self-attention was used to refine temporal and intra-modal features, leveraging long-range interactions to enhance emotion recognition performance.

Zhang \emph{et al.} \cite{Zhang2024} used multimodal Transformers for integrating EEG and peripheral physiological signals for valence and arousal classification tasks. The fusion methodology employed an intra-modality self-attention to enhance modality-specific features, followed by an inter-modality cross-attention to model correlations between modalities. A credibility fusion module then evaluated the sequential pattern consistency of modalities using dynamic time warping, assigning weights to ensure credible fusion.

Zhu \emph{et al.} \cite{Zhu2025} employed EEG and eye-tracking data for mild depression detection. The fusion was achieved through multimodal Transformers under hierarchical attention to model temporal dependencies and inter-modal interactions. The model combined intra-modal feature extraction with inter-modal dependencies in a structured intermediate fusion framework.

\textbf{Challenges in Emotion Labeling and Elicitation} The future of building effective affect decoding systems lies in addressing fundamental limitations in current datasets. One critical issue is the labeling process of emotions, which is a comparatively more serious problem for brain data which is hard to annotate directly by human evaluators. Emotion labeling also involves subjectivity and cultural diversity involved in its definition and expression. Current datasets often rely on self-reported labels or annotations from multiple experts, while future efforts could also leverage AI techniques such as self-supervised and semi-supervised learning to reduce the reliance on labels beyond incorporating multiple modalities for better performance.

Another significant challenge is the reliance on video stimuli for emotion elicitation, which may fail to accurately reflect real-world emotional experiences. Videos often evoke only superficial emotional responses or limited classes of labels. Moving forward, affective BCIs would benefit from the development of more intuitive, naturalistic, and interactive emotion-elicitation paradigms.

\section{Progress, Prospects, and Pitfalls} \label{sect:prospects}

\subsection{Multimodal Big Data} \label{sect:bigdata}

The advancement of multimodal BCIs depends on the availability of large-scale, high-quality datasets. AI-driven decoding methodologies have significantly expanded the scope of insights extractable from such data. For multimodal BCIs, such data must not only be extensive but also meticulously structured to suit specific algorithmic tasks.

\textbf{Importance of Big Data} AI techniques thrive on large datasets to uncover complex, non-linear relationships within and across modalities for training heavily parameterized neural network models. For instance, sequential models like Transformers or recurrent neural networks require extensive temporal data to learn long-range dependencies, while generative models depend on substantial datasets to accurately align and reconstruct multimodal representations. Consequently, the size of multimodal datasets must scale to ensure models are adequately trained and generalizable.

Large-scale initiatives like the Human Connectome Project \cite{VanEssen2012, Glasser2013} exemplify the importance of comprehensive datasets, offering thousands of fMRI scans enriched with supplementary information such as behavioral and genetic data. However, the scale of data required for multimodal BCIs often extends beyond the capabilities of such traditional repositories, necessitating new approaches to data acquisition, annotation, and integration.

\textbf{Data that are Paired and Aligned} Mapping or translation tasks involve converting data from one modality to another. For mapping tasks, datasets must consist of paired data from two modalities, where each instance in one modality directly corresponds to an instance in another. For sequential translation tasks, the alignment must also be temporally correlated. Pairing and alignment are the core requirements for such settings that enable AI algorithms to function effectively \cite{Sato2024}.

\textbf{Data of Complementary Modalities} Fusion tasks aim to integrate data from multiple modalities to enhance decoding accuracy or robustness. Unlike mapping/translation tasks, fusion tasks focus on combining complementary information from different sensors. For such tasks, simultaneous data collection is crucial to ensure that signals across modalities reflect the same underlying cognitive or physiological state. Note that complementarity of sensor modalities is the key to fusion. Modalities should provide distinct but related perspectives of the same phenomenon. For instance, EEG offers high temporal resolution, while fMRI provides detailed spatial information. AI-driven fusion models rely on the availability of synchronized datasets to effectively combine these diverse perspectives.

\subsection{Brain Data Heterogeneity} \label{sect:heterogeneity}

\begin{figure*}[htbp]\centering
\includegraphics[width=\linewidth,clip]{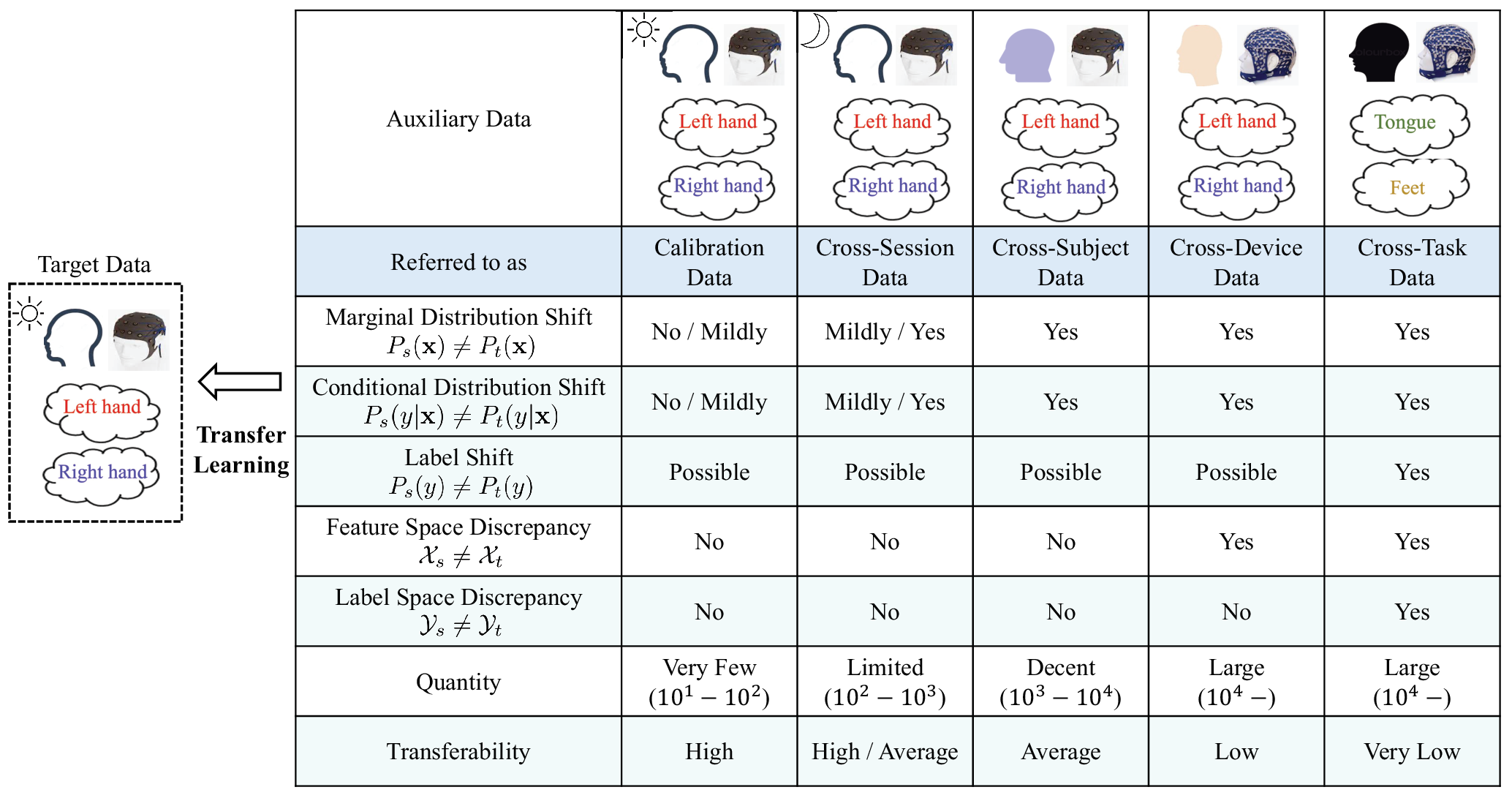}
\caption{Available types of auxiliary data that could benefit target classification, along with their types of shifts, general quantity, and availability, from a TL perspective.} \label{fig:transfer}
\end{figure*}

A significant obstacle in brain data decoding is the limited dataset size available for machine learning. Instead of laboriously collecting new data, a promising alternative enabled by AI algorithms is transfer learning \cite{Pan2010, Zhuang2021}. Transfer learning leverages auxiliary labeled data previously collected, presumably from additional subjects acquired with the same or even a different brain data collection device, thereby improving the decoding accuracy for the target subject~\cite{Wu2022a, Wu2022}. Most studies focus on auxiliary data from subjects under identical collection devices and tasks. However, cross-device and cross-task transfers, where brain data are gathered using different devices or under varying tasks, remain underexplored. Fig.~\ref{fig:transfer} summarizes the possible types of auxiliary data, along with their relationships to the test data, from a transfer learning perspective.

\textbf{Inter-Subject Discrepancy} Brain data (and physiological data in general) varies significantly across subjects/users~\cite{Wu2022a}. Without specific adaptation strategies, training AI algorithms with auxiliary subject data often yields poor or even negative results for the target subject~\cite{Zhang2023}. These variations can be interpreted as distribution shifts in a probabilistic sense~\cite{Pan2010, Zhuang2021}, including marginal distribution shifts, conditional distribution shifts, and label shifts, which may occur independently or simultaneously, necessitating distinct mitigation strategies.

The literature provides abundant solutions to address these challenges. Data normalization is the primary method for handling marginal distribution shifts~\cite{Apicella2023}, with data alignment techniques offering effective solutions~\cite{Zanini2018RA, He2020EA, Wu2022}. More complex issues like conditional distribution or label shifts demand fine-grained approaches \cite{Wong2020, Zhang2020}. Recent works have explored sophisticated and robust learning objectives to mitigate these impacts particularly for neural network models \cite{Li2024T-TIME, Wang2022, Li2020b, Wang2023c}. Additionally, studies consistently report that brain data shifts not only exist across subjects but also occur within the same subject across different collection sessions~\cite{Wu2022, Li2024T-TIME}.

\textbf{Heterogeneity in Collection Devices} Unlike image or text data, which can be easily aggregated for deep learning, brain data such as EEG or other intracranial multi-electrode electrical signals are complex, multivariate time-series data. These signals are collected using varying numbers of electrodes, electrode placements, and trial durations. Most existing algorithms rely on single-channel feature extraction or neural networks designed for fixed input dimensions. Aligning heterogeneous EEG data from diverse devices is challenging but essential for building large-scale datasets and enabling advanced AI applications.

Device heterogeneity has been addressed in limited studies. Some approaches utilize a common subset of channels across training and testing datasets~\cite{Xu2020a, Xie2023}. Nakanishi \emph{et al.}~\cite{Nakanishi2020} proposed mitigating device discrepancies using spatial filters computed via channel averaging. However, such methods often result in information loss by failing to utilize the full electrode set.

Liu \emph{et al.} \cite{Liu2024a} proposed a co-training approach utilizing hand-designed network architectures to facilitate feature space alignment for motor imagery EEG classification. Such an approach requires specifically designed filters for each device. Li \emph{et al.} \cite{Li2025HSDA}  proposed a unified framework of heterogeneous supervised domain adaptation to address feature space discrepancies along with marginal and conditional distribution shifts simultaneously. Their approach uses symmetric heterogeneous channel alignment to align channels across devices. Additionally, they introduced a distance metric-based feature learning objective and a model inversion-based gradient regularization strategy to enhance class discriminability while preserving transferability.

An alternative to symmetric alignment is an asymmetric, one-way mapping function from source to target devices. Liu \emph{et al.} \cite{Liu2025} proposed a knowledge distillation-based approach for motor imagery EEG classification, ensuring information transfer from full channels of the source device to the common subset of channels shared by both the source and target device. Wang \emph{et al.} \cite{Wang2025} proposed to further transfer across species (canines and humans) for EEG-based epileptic seizure detection, using knowledge distillation to match source and target channels in the representation space.

\textbf{Different Task Transfer} Labeled data from similar or related tasks can also facilitate the decoding of a new task~\cite{Wu2022a}. For instance, motor imagery EEG data for foot movement can aid the classification of left- and right-hand movements, a scenario referred to as different task transfer~\cite{He2020LA}. In such cases, the source and target domains have different label spaces (tasks). A transformation matrix, constructed using a small amount of labeled target data, can project and align the target distribution with the source distribution for the most relevant class. Such utilization of auxiliary data from different tasks has also demonstrated effectiveness in practice \cite{Vishwanath2022}.

\textbf{Challenges for Transfer Learning} Despite its promise, transfer learning remains constrained by the limitations of classic deep learning, including:
\begin{itemize}
\item Dependence on Distribution Estimation: Transfer learning performance often relies on the availability of source and target domain data, with significant degradation observed when such data are limited, inaccessible, or under poor quality.
\item Limited Generalization: Current methods struggle to adapt across diverse paradigms, reducing their applicability in real-world settings.
\item Single-Modality Focus: Most transfer learning frameworks are extensively theorized, studied, and validated within a single modality, limiting their applicability in multimodal contexts.
\end{itemize}
The emerging paradigm of the brain foundation model, discussed in the next subsection, offers a potential revolutionary alternative to address these issues, representing a promising direction for future research.

\subsection{Brain Foundation Model} \label{sect:foundation}

The trend toward building larger and deeper models is evident across fields, from natural language processing \cite{Brown2020} to computer vision \cite{Radford2021}, and is now extending to general time-series data \cite{Wang2024a}. Developing large brain foundation models has gained significant interest in the research community.

\textbf{Large-scale Transformers}  In the context of brain data, the advantages of Transformer architectures are being increasingly recognized. Architectures that integrate CNN-based feature extraction with Transformer encoders have demonstrated superior performance in brain data decoding when compared to traditional CNN- or RNN-only models \cite{Schirrmeister2017, Lawhern2018EEGNet, Sakhavi2018, Song2020, Ding2023, Ding2024, Ding2024a}. These hybrid designs, when properly optimized, offer notable improvements in decoding accuracy.

\textbf{Training Strategies Fit for Foundation Models} Scaling both model size and data volume is necessary but insufficient for building effective foundation models. Tailored training strategies that leverage large datasets are critical. A key component of these strategies is representation learning, often achieved through self-supervised learning \cite{Liu2023b}. This approach eliminates the dependence on labels or task-specific settings, enabling the aggregation of diverse datasets. The objective is to learn generalized features from brain data that can be universally applied to various downstream tasks.

\textbf{Tokenization for Embedding} Tokenization is a crucial challenge in applying Transformer models to multivariate time-series analysis. For brain signal modeling, tokenization can be implemented in multiple ways, such as using single channels \cite{Zhang2022a, Yang2023}, patches of time windows \cite{Zhang2023Brant, Wang2024b}, combinations of channel subsets and time windows \cite{Jiang2024LaBraM}, or frequency-domain representations \cite{Wang2023BrainBERT}. In multimodal learning, tokenization becomes even more complex, requiring a unified strategy to accommodate multiple modalities within a single framework.

In the context of multimodal applications, research interest has also focused on generative brain models \cite{Mai2024} and aligning brain data with other modalities \cite{Xia2024a}. However, the development of multimodal brain foundation models remains an open challenge, with key questions yet to be resolved.

\subsection{Common Errors} \label{sect:pitfall}

Physiological signals often exhibit temporal correlations within both short and long ranges during experimental sessions, as demonstrated in EEG \cite{LinkenkaerHansen2001} and fMRI \cite{Bullmore2001}. These temporal patterns indicate the presence of a background hidden state unique to each session or block of data collection. Such hidden states, although generally unrelated to the task stimuli, are inherent in data collection processes. Consequently, machine learning algorithms should not rely on these hidden states as proxies for task-related classes or exploit future hidden states to predict past outcomes. Due to the interdisciplinary nature of brain signal decoding, these issues might not always be evident to researchers, particularly those in the machine-learning community focused on multimodal applications. Below, we examine some of the most notable problems in this context.

\begin{figure*}[htpb]
\centering
\subfigure[]{\includegraphics[width=.55\linewidth,clip]{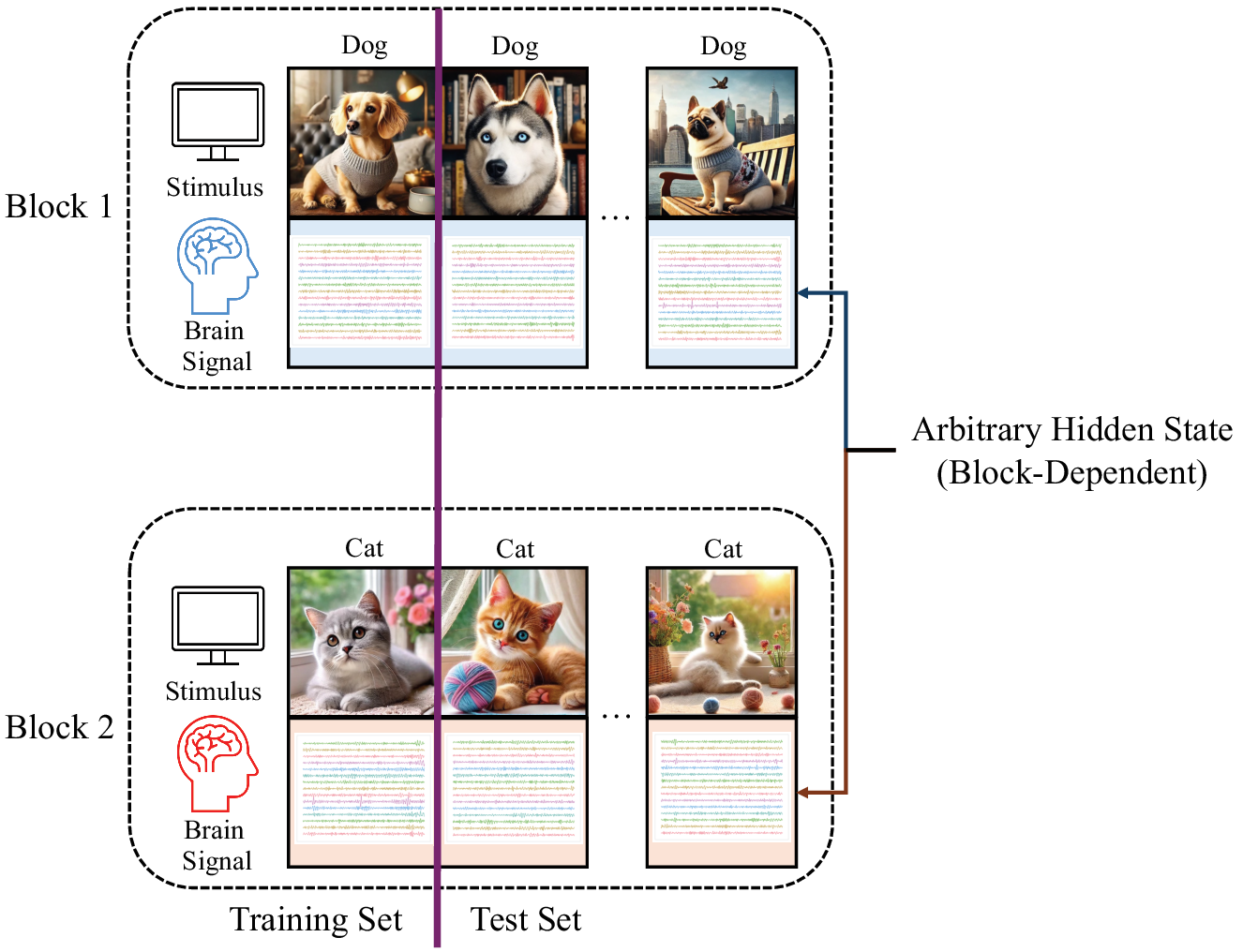}\label{fig:pitfall_wrong}}
\subfigure[]{\includegraphics[width=.39\linewidth,clip]{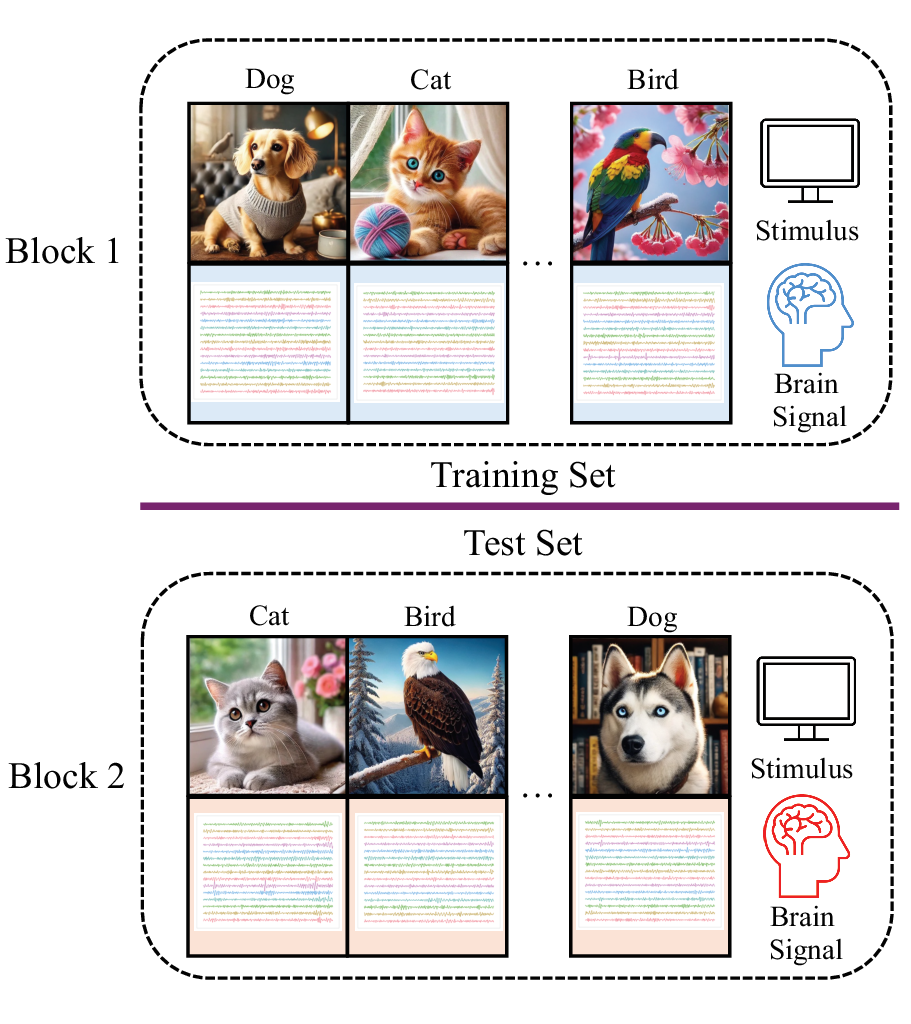}\label{fig:pitfall_correct}}
\caption{Common error of block design on brain signal classification experiments. (a) Incorrect experiment protocol, where each block contains only one class of stimuli. Classifiers can rely solely on the block-specific brain states to differentiate between task-related classes, irrespective of the actual stimulus-related brain patterns. This limitation persists regardless of the random or sequential division between training and test sets. (b) The correct experiment protocol, where each block would contain all classes of stimuli and the stimuli would be preferably presented in random order. During analysis, training and test samples are segregated to not overlap by block. It ensures that classifiers do not learn to discriminate based merely on block-specific temporal patterns.} \label{fig:pitfall}
\end{figure*}

\textbf{Block Design} A recent study \cite{Li2021} highlights errors in certain data collection protocols, such as those reported in \cite{Spampinato2017}, and subsequent works \cite{Palazzo2017, Kavasidis2017, Du2018, Kumar2018, Tirupattur2018, Palazzo2021}. These errors stem from block-wise temporal correlations inherent in brain signals, leading to unintended biases. The study illustrates how block design flaws, as shown in Fig.~\ref{fig:pitfall}, can cause classification algorithms to rely on block-specific hidden states rather than task-relevant patterns. This issue arises when training and test splits fail to ensure temporal independence. As a result, regardless of whether the classifier is feature-based or neural-network-based, its predictions are inadvertently guided by block-specific temporal artifacts rather than stimulus-related activity. Consequently, performance metrics appear artificially inflated and fail to reflect real-world applicability. Specifically, \textit{``The block design leads to the classification of arbitrary brain states based on block-level temporal correlations that are known to exist in all EEG data, rather than stimulus-related activity. Because every trial in their test sets comes from the same block as many trials in the corresponding training sets, their block design thus leads to classifying arbitrary temporal artifacts of the data instead of stimulus-related activity."}

\textbf{Temporal Normalization} Brain-state monitoring spans durations from seconds to months, depending on the clinical context or task stimuli \cite{Gilron2021}. For brain signals with low signal-to-noise ratios and susceptibility to motion artifacts, normalization techniques are essential. These approaches, commonly considered preprocessing steps in machine learning, are increasingly recognized in brain-signal analysis. Temporal normalization addresses marginal distribution shifts \cite{Wu2022a, Apicella2023} by reducing noise and variability, enhancing decoding performance. However, when applied without caution, especially in the presence of block design flaws, normalization can introduce temporal leakage, compromising model validity.

\begin{figure}[htpb] \centering
\includegraphics[width=\linewidth,clip]{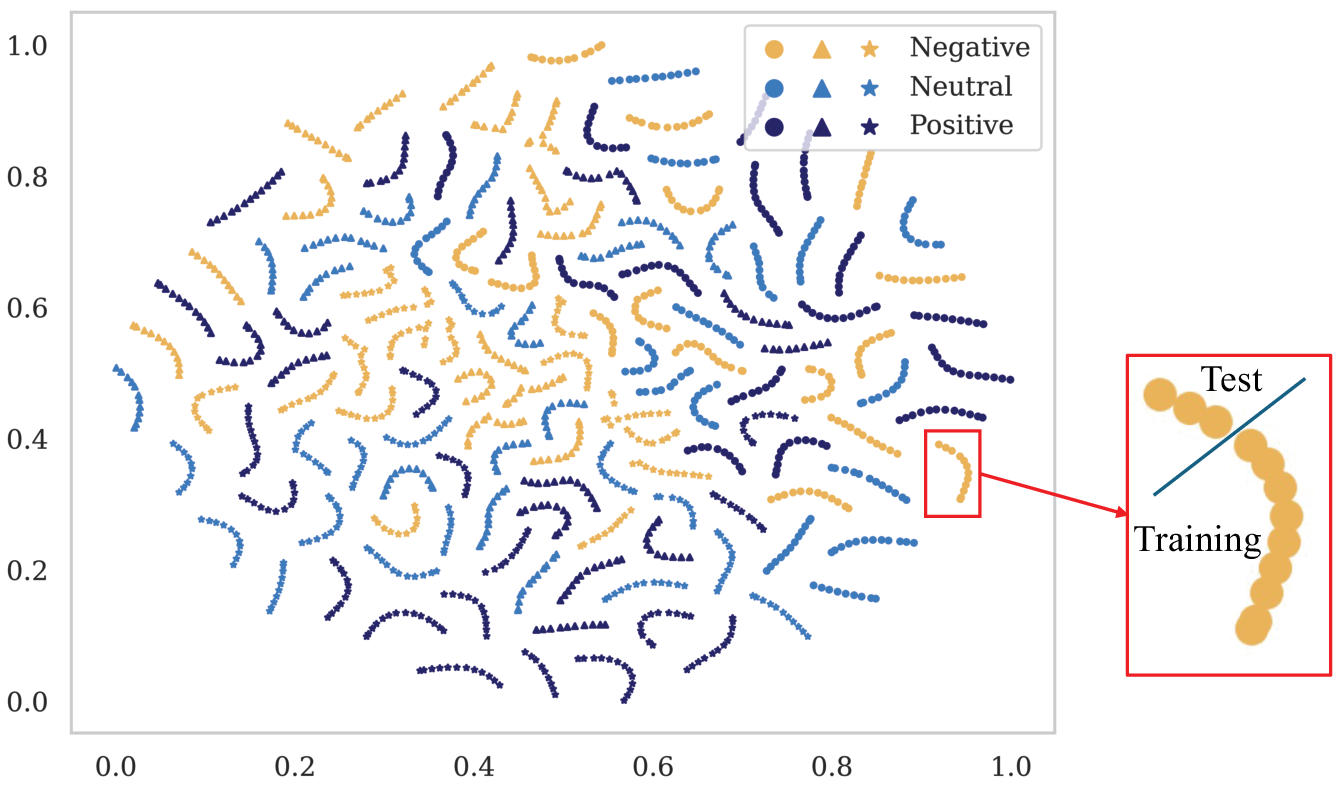}
\caption{Common error of normalization on block-based brain signal classification under machine learning. $t$-SNE visualization of EEG data from SEED dataset shows that normalization produces high within-block feature similarity while maintaining distinct distributions across blocks. Improper data splits involving the same block in training and test sets lead to misleading performance metrics.} \label{fig:norm}
\end{figure}

For instance, the SEED dataset \cite{Zheng2015}, widely used in affective BCIs, applies differential entropy features normalized via linear dynamical systems \cite{Shi2010}. In this context, users' emotional states are analyzed based on brain signal recordings elicited by movie clips. During analysis, each multi-minute recording block is typically split into smaller trials. Normalization is applied to these trials using statistics from preceding trials within the same block. As illustrated in Fig.~\ref{fig:norm}, this leads to block-specific feature distributions. When training and test sets are drawn from the same block (thus containing highly similar normalized features of the identical class), classifiers exploit these distributions, resulting in inflated performance metrics. Proper segregation of training and test samples across blocks is crucial to avoid this issue.

\textbf{Avoiding Temporal Leakage} To highlight these issues, we emphasize the following:
\begin{itemize}
\item Data Split Protocol: Temporal leakage occurs when training and test sets include trials from the same block, and classifiers rely on brain states instead of task-related components for classification.
\item Classifier Independence: The core issue stems from improper data splits combined with flawed experimental protocols. Temporal leakage is independent of the classifier type.
\item Impact on Experimental Conclusions: Temporal leakage, of false data split under such methods, invalidates experimental conclusions for decoding. However, it does not necessarily falsify the classification algorithm (neural network architecture or feature extraction method) or the normalization method itself.
\end{itemize}

Researchers can adopt the following practices to avoid temporal leakage:
\begin{itemize}
\item Adopt Standard Experimental Designs: Ensure experimental conditions are randomly distributed across multiple blocks, intertwining task stimuli with block-specific hidden states.
\item Avoid Random Data Splits: Random splits of brain data within a block into training and test sets are impractical and incur temporal leakage, compromising real-world applicability \cite{Kapoor2023}.
\item Focus on Cross-Subject Generalization: Test trials should ideally come from entirely separate blocks or subjects to ensure robust model evaluation, regardless of whether it is feature-based models, neural network models, or foundation models.
\item Use Proper Preprocessing: While preprocessing steps (e.g., bandpass, notch filtering) cannot entirely eliminate the impact of flawed experimental designs, they can mitigate issues or reveal underlying problems \cite{Li2021}.
\end{itemize}

\subsection{Security and Privacy} \label{sect:securityandprivacy}

The security and privacy of BCIs have long been major concerns \cite{Ienca2018}. As multimodal techniques and large models continue to advance, the significance of BCI security and privacy will only increase \cite{Turchet2024, Magee2024}.

\textbf{Security Issues} BCI outputs are susceptible to manipulation via adversarial attacks \cite{Zhang2021a}. Zhang and Wu \cite{Zhang2019} were the first to demonstrate that adversarial samples could significantly degrade the decoding accuracy of EEG-based BCIs. Liu \emph{et al.} \cite{Liu2021} and Jung \emph{et al.} \cite{Jung2023} developed universal adversarial perturbations, simplifying the implementation of such attacks. Wang \emph{et al.} \cite{Wang2022a} explored physically constrained attacks, while Gunawardena \emph{et al.} \cite{Gunawardena2024} showed that attacking a single sensor could manipulate the outputs of smart headsets for behavioral biometrics. Liang \emph{et al.} \cite{Liang2024} discovered that backdoors could be embedded in multimodal models, allowing for manipulation of their outputs. These adversarial attacks raise significant safety concerns, making adversarial robustness a critical issue for BCIs.

Both active and reactive defense strategies are essential for enhancing BCI security \cite{Gunawardena2024}. Active defense can be integrated at various stages of the BCI decoding pipeline, with approaches such as standardized data processing, effective data augmentation, rational model architectures, and generalizable training algorithms contributing to BCI models that are more resilient to adversarial attacks \cite{Chen2024ABAT, Croce2021}. Reactive defense focuses on detecting and rejecting inputs that have been tampered with. Combining both defense strategies can improve the overall security of BCIs \cite{Gunawardena2024}.

As multimodal techniques and large models evolve, new security challenges must be addressed. For example, the integration of multimodal, cross-paradigm brain signals may influence the vulnerability of BCIs, an area yet to be explored \cite{Liang2024}. This could lead to the emergence of new methods to enhance BCI security, such as improved data alignment in multimodal models and more effective multi-dataset fusion in large models, which may help create more secure and accurate BCI models. Consequently, adversarial robustness could also serve as an evaluation metric for BCI models \cite{Meng2023}, facilitating the development of more accurate and secure BCI models.

\textbf{Privacy Issues} Several regulations, such as the European Union's General Data Protection Regulation and China's Personal Information Protection Law, have been introduced to safeguard user privacy. Regardless of the paradigm, brain data inherently contain sensitive personal information \cite{Xia2023}. For example, Martinovic \emph{et al.} \cite{Martinovic2012} showed that EEG signals can reveal private details, including credit card numbers, personal identification numbers, contacts, and addresses. Ienca \emph{et al.} \cite{Ienca2018} highlighted the ethical and privacy issues associated with consumer neurotechnology. Landau \emph{et al.} \cite{Landau2020} demonstrated that even resting-state EEG data can disclose personal characteristics such as personality traits and cognitive abilities.

Various methods can be employed to protect the privacy of brain data, including privacy-preserving machine learning, synthetic data generation, data perturbation, and federated learning \cite{Chen2024UWP, Xia2023}. Privacy-preserving machine learning avoids using raw brain data or model parameters directly. Synthetic data generation leverages generative models to produce data that retains essential information for BCI tasks. Data perturbation adds noise to the original data to protect private information, while maintaining the data's utility for downstream tasks \cite{Chen2024UWP}. Federated learning trains a global model without sharing brain data between the server and the clients, or among the clients, protecting user privacy by preventing other devices from accessing raw data stored on the local client, thus avoiding the privacy risks of centralized datasets \cite{Jia2024, Liu2024a}.

As more brain signals are incorporated into multimodal and large BCI decoding models, ensuring privacy will present greater challenges. Future research must focus on understanding the privacy risks in larger datasets and protecting user information.

\section{Conclusions} \label{sect:conclusions}

This review explored the landscape of AI-powered decoding methodologies for multimodal BCIs, addressing their elements, decoding algorithms, applications, and challenges. By emphasizing advancements in cross-modality mapping, sequential modeling, and multimodal fusion, this work highlighted the pivotal role of AI decoding algorithms in enabling more accurate BCIs. From decoding brain signals for visual, speech, and affective BCI applications to discussing the transformative potential of large-scale brain foundation models, we showcased the profound impact of AI methodologies for this interdisciplinary field.

The prospects for AI-powered multimodal BCIs are vast, extending into healthcare, neurorehabilitation, immersive virtual reality, and beyond. With continued advancements in decoding algorithms and the integration of large-scale brain data, these systems are poised to transform human-computer interaction and neurotechnology. By addressing the outlined challenges and leveraging emerging AI paradigms, the field is well-positioned to unlock the full potential of BCIs. Fundings also support the development of BCIs \cite{Wang2017, Miller2024BRAINInit}, paving the way for a future where seamless brain-machine integration enhances human capabilities and quality of life.

\bibliographystyle{IEEEtran} \bibliography{mmbci}

\end{document}